\documentclass[aps,prd,twocolumn,superscriptaddress,nofootinbib,floatfix]{revtex4-2}
\usepackage{graphicx}
\usepackage{amsmath,amssymb}
\usepackage{bm}
\usepackage{placeins}
\usepackage{xcolor}
\usepackage[colorlinks=true,linkcolor=blue,citecolor=blue,urlcolor=blue]{hyperref}

\newcommand{\discrho}{1.91}

\newcommand{\discphi}{1.44}

\newcommand{\normrhoSOne}{0.93}
\newcommand{\normSpreadrhoSOne}{0.17}
\newcommand{\normrhoSTwo}{0.57}
\newcommand{\normSpreadrhoSTwo}{0.08}

\newcommand{\chiTwoSOne}{2.03}
\newcommand{\chiTwoSOneRaw}{69}
\newcommand{\chiTwoSTwo}{8.57}
\newcommand{\chiTwoSTwoRaw}{291}

\newcommand{\nDatasets}{75}
\newcommand{\nPoints}{442}

\newcommand{\sigmaZero}{58.9}
\newcommand{\gammaS}{0.740}
\newcommand{\lambdaCGC}{0.161}
\newcommand{\xZero}{1.16\times10^{-7}}
\newcommand{\Nzero}{0.7}
\newcommand{\kappaZero}{9.9}
\newcommand{\omegaEps}{0.98}
\newcommand{\cgcChiN}{1719/612=2.81}

\begin{document}

\title{Diffractive Vector-Meson Production from the Light-Front Quark Model}

\author{Satyajit Puhan}
\email{puhansatyajit@gmail.com}
\affiliation{Institute of Physics, Academia Sinica, Taipei 11529, Taiwan}

\date{\today}

\begin{abstract}
We study exclusive diffractive electroproduction of the $\rho^0$, $\phi$, $J/\psi$, and $\psi(2S)$ mesons in the color-dipole picture using two spin-orbit light-front wave functions, S-1 and S-2. S-1 follows from the Melosh--Wigner rotation in the light-front quark model (LFQM), while S-2 is the spin-improved form commonly used in holographic and boosted-Gaussian approaches. Both share the same radial wave function, with their parameters determined from the LFQM Hamiltonian. The color-glass-condensate dipole parameters are fixed from inclusive structure-function data, with no fit to diffractive data. We compare the predictions with measurements of cross sections, $R=\sigma_L/\sigma_T$, $t$ and $W$ dependences, and $\sigma_\phi/\sigma_\rho$. S-1 gives a better description of the light-meson observables, while S-2 provides a better description of the charmonium ratio $\sigma_{\psi(2S)}/\sigma_{J/\psi}$ and the $J/\psi$ decay constant. We also present the charge, magnetic, and quadrupole form factors of the four mesons and compare them with available lattice-QCD results.
\end{abstract}

\maketitle

%======================================================================
\section{Introduction}
\label{sec:intro}
%======================================================================

Exclusive electroproduction of a vector meson, $\gamma^*p\to Vp$, is among the cleanest
probes of the proton at small Bjorken $x$. The proton stays intact, the final state is a single
meson, and at high energy the process is driven by the exchange of a colorless pair of gluons, so
that the cross section responds to the square of the gluon density rather than to the density
itself \cite{Brodsky:1994kf,Frankfurt:1995jw,Martin:1996bp,Ivanov:2004ax}. H1 and ZEUS measured
light vector-meson production over a wide range of photon virtuality $Q^2$, energy $W$ and
momentum transfer $t$ \cite{H1rho1996,ZEUSrho1999,H1rho2000,ZEUSphi2005,ZEUSrho2007,H1rho2010},
as well as $J/\psi$ and $\psi(2S)$ production
\cite{ZEUS:2004jpsi,H1:2005jpsi,H1:2013jpsi,H1:1999psi,H1:2002psi,ZEUS:2016psi,ZEUS:2022psi}.
The fixed-target experiments NMC, E665 and HERMES added data at lower energies
\cite{NMCrho1994,E665rho1997,HERMESrho2009}, and the process is a key part of the program of the
future Electron-Ion Collider (EIC) \cite{AccardiEIC,EICYellowReport}, where it will be used to
image the gluons in the transverse plane and to look for saturation.

\par Future facilities will measure these observables with much better precision. At the EIC,
with $\sqrt s\approx20$--140~GeV \cite{AccardiEIC,EICYellowReport}, $\rho^0$, $\phi$, $J/\psi$
and $\psi(2S)$ production will be measured between the fixed-target and HERA energies, and
$\sigma_L$ and $\sigma_T$ can be separated by varying the beam energies, which gives $R$
directly. The EicC, with $\sqrt s\approx15$--20~GeV \cite{EicC:2021}, will study charmonium
production close to threshold, and the LHeC, with $\sqrt s\approx1.2$~TeV \cite{LHeC:2020}, will
reach much smaller $x$.

\par Several frameworks are used to describe these data
\cite{Shi:2025mne,Gurjar:2024wpq,ForshawSandapen2012,AhmadyEtAl2016,Gurjar:2025kcp,Mantysaari:2022kdm,Lappi:2020ufv}.
In collinear factorization, the amplitude for a longitudinal photon is expressed through
generalized parton distributions, and the handbag approach of Goloskokov and Kroll extends this
to transverse photons by keeping the transverse momentum of the partons
\cite{Goloskokov:2005sd,Goloskokov:2007nt}. At small $x$, the color-dipole picture
\cite{Nikolaev:1990ja,Mueller:1993rr} is especially convenient. Seen from the proton rest frame,
the virtual photon splits into a quark--antiquark pair long before the interaction. The pair
scatters elastically off the proton as a color dipole of fixed transverse size and then forms the
vector meson. The amplitude therefore factorizes into the photon wave function, which is known
from QED, the dipole--proton cross section, which is fitted to inclusive deep-inelastic data
\cite{Golec-Biernat:1998zce,IancuItakuraMunier2004,Kowalski:2003hm,Rezaeian2013}, and the
light-front wave function (LFWF) of the meson. The approach has been applied to light vector
mesons for many years \cite{Nemchik:1996cw,KowalskiMotykaWatt2006,Forshaw:2003ki}, and it
describes the HERA data well once a suitable meson wave function is chosen
\cite{ForshawSandapen2012,AhmadyEtAl2016,Armesto:2014sma,Gurjar:2024wpq}.

\par The meson LFWF cannot be taken from other processes and has to be modeled. It is the product
of a radial function and a spin-orbit function, the latter fixing how the quark and antiquark
helicities combine into the helicity of the meson. The radial part is usually a boosted Gaussian
\cite{Nemchik:1996cw,Forshaw:2003ki,KowalskiMotykaWatt2006} or the wave function of
light-front holography \cite{Brodsky:2014yha,ForshawSandapen2012}, and the differences between
the two are well studied. The spin-orbit part has received far less attention. In the
light-front quark model (LFQM) \cite{ChoiJi1999,ChoiJi2007,Arifi:2025olq,Arifi:2022qnd} it follows
from the Melosh--Wigner rotation \cite{Melosh1974} of the rest-frame spin state of the pair, a
relativistic effect that depends on the invariant mass of the pair. The dipole studies use instead
a simpler, ``spin-improved'' vertex whose longitudinal component contains the meson mass $M_V$
\cite{ForshawSandapen2012,AhmadyEtAl2016,Gurjar:2024wpq}. The two forms agree in the
nonrelativistic limit but not for light quarks. The LFQM wave function has been used extensively
for the valence structure of the $\rho$ meson \cite{Tanisha:2025qda,Acharyya:2024enp}, and in a
recent study, we found that the choice of spin structure changes the parton distributions of the
$\rho$ considerably, in particular the tensor-polarized ones \cite{Puhan:2025ujg}. Here, we use
both spin wave functions and compare the resulting LFQM predictions.

\par Absolute cross sections alone cannot separate the two spin-orbit wave functions. Besides the
meson wave function, they depend on the normalization of the dipole cross section, on the
treatment of gluon skewedness \cite{Shuvaev:1999ce} and on the parametrization of the diffractive
slope, and each of these can move the result by as much as the difference between S-1 and S-2.
Observables in which these factors drop out are therefore needed. The ratio of the longitudinal to
the transverse cross section, $R=\sigma_L/\sigma_T$, is one of them. It is extracted from the
decay angular distribution through the spin-density matrix elements \cite{Schilling:1973ag}, all
normalization factors cancel in it, and it measures directly the relative weight of the
helicity-zero and helicity-one components of the meson wave function. We concentrate on $R$ and
on the other quantities free of this normalization, namely the ratio $\sigma_\phi/\sigma_\rho$,
the decay constants, which do not involve the dipole cross section at all, and the
electromagnetic form factors, which test the same wave functions in a different process and can be
compared with lattice QCD (LQCD) \cite{QCDSF:2008tjq,Shultz:2015pfa}. For the form factors, we also examine the
light-front angular condition~\cite{Grach:1983hd,Brodsky:1992px,Choi:2004ww}, which shows how well
each spin structure preserves the rotational covariance of the current in the truncated $q\bar q$
Fock space.

\par For the $J/\psi$ and $\psi(2S)$, the Melosh--Wigner rotation matters less than for the light
mesons, but the radially excited $\psi(2S)$ is sensitive to the shape of the radial wave function.
H1 and ZEUS measured the ratio of the $\psi(2S)$ and $J/\psi$ cross sections in photoproduction
and in deep-inelastic scattering as a function of $Q^2$ and $W$
\cite{H1:1998psi,H1:1999psi,H1:2002psi,ZEUS:2016psi,ZEUS:2022psi}, and the same ratio has recently
been measured in ultraperipheral collisions at the LHC \cite{LHCb:2024psi}. It is well below one
because the $\psi(2S)$ wave function has a node, so that dipoles smaller and larger than the node
contribute with opposite signs \cite{Nemchik:1997xb,Hoyer:1999xe,Suzuki:2000az,Frankfurt:1997fj}.
Like $R$, the ratio does not depend on the normalization of the dipole cross section, and it has
been studied with a variety of wave functions and dipole models
\cite{Sharma:2023njj,KowalskiMotykaWatt2006,Cepila:2019skb,Krelina:2019gee,Krelina:2019egg,Henkels:2020kju,Lappi:2020ufv,Mantysaari:2021ryb,Peredo:2023oym,Jamal:2026heavy}.
In the LFQM, radially excited states are built from the harmonic-oscillator basis in refs. 
\cite{Dhiman:2019ddr,Arifi:2022pal,Ridwan:2025}. Lattice QCD results are also available for the
$J/\psi$ form factors \cite{Dudek:2006ej,Delaney:2023fsx,Li:2018jpsi}.

\par In this paper, we calculate the measured diffractive observables of the $\rho^0$ and $\phi$ mesons, as well as the ratio $\sigma_{\psi(2S)}/\sigma_{J/\psi}$, using both spin structures and compare our results with the available experimental data. The input parameters, including the quark masses and the harmonic scale parameter, are determined by solving the two-body Hamiltonian using the standard variational principle. For the light mesons, we employ a harmonic confining potential following Ref.~\cite{Choi:1997iq}, while for the heavy charmonia, we use the screened confining potential of Ref.~\cite{Ridwan:2024ngc}, which incorporates the screening effect. The quark masses enter the photon as well as the meson wave functions, and
we determine the parameters of the dipole cross section by fitting the structure-function data of
H1 and ZEUS, E665 and NMC with the same quark masses. Apart from the spin structure, every input is the
same in the two calculations and no parameter is adjusted to the diffractive data, so any
difference between the results comes from the spin structure alone. The decay constants and the
electromagnetic form factors of the four mesons are calculated with the same wave functions.

\par The paper is organized as follows. Section~\ref{sec:model} introduces the two LFWFs.
Section~\ref{sec:diffractive} describes the dipole framework and compares the diffractive cross
sections, including $\sigma_{\psi(2S)}/\sigma_{J/\psi}$, with the data. The decay constants, the
electromagnetic form factors and the angular condition of the four mesons are discussed in
Sec.~\ref{sec:emff}, and Sec.~\ref{sec:summary} summarizes our results. The normalization of the
S-2 wave function, the variational determination of the LFQM parameters, and the fit of the
dipole cross section are given in Appendices~\ref{app:norm}, \ref{app:var} and \ref{app:cgc}.

%======================================================================
\section{Vector-meson light-front wave function}
\label{sec:model}
%======================================================================

In light-front quantization a meson is expanded in Fock states of quarks,
antiquarks and gluons as~\cite{Brodsky:1997de,LepageBrodsky1980}
\begin{equation}
\begin{split}
|V\rangle=&\sum|q\bar q\rangle\,\Psi_{q\bar q}
+\sum|q\bar q g\rangle\,\Psi_{q\bar q g}\\
&+\sum|q\bar q q\bar q\rangle\,\Psi_{q\bar q q\bar q}+\cdots .
\end{split}
\label{eq:fock}
\end{equation}
In the LFQM, only the lowest, quark--antiquark Fock state is kept, and the
effect of the higher states is absorbed into the constituent quark mass. The
state of a vector meson with momentum $P=(P^+,P^-,\mathbf{P}_\perp)$ and
polarizations $\Lambda=(\pm 1,0)$ is expressed as~\cite{Puhan:2025ujg,Tanisha:2025qda}
\begin{align}
|V(P,\Lambda)\rangle={}&\sum_{h,\bar h}\int
\frac{dx\,d^2\mathbf{k}_\perp}{\sqrt{x(1-x)}\,2(2\pi)^3}\,
\Psi^{\Lambda}_{h\bar h}(x,\mathbf{k}_\perp)\nonumber\\
&\times\big|xP^+,\mathbf{k}_\perp,h;\,(1-x)P^+,-\mathbf{k}_\perp,\bar h\big\rangle ,
\label{eq:twobody}
\end{align}
where, $x$ and $1-x$ are the longitudinal momentum fractions of the quark and
the antiquark, $\mathbf{k}_\perp$ is their relative transverse momentum, and
$h$ and $\bar h$ are their helicities, respectively. The LFWF does not depend on the meson
momentum $P$, so it is boost invariant, and it is written as the product of a
radial $\phi(x,\mathbf{k}_\perp)$ and a spin-orbit $\mathcal{S}^{\Lambda}_{h\bar h}(x,\mathbf{k}_\perp)$ part as
\begin{equation}
\Psi^{\Lambda}_{h\bar h}(x,\mathbf{k}_\perp)
  = \phi(x,\mathbf{k}_\perp)\,\mathcal{S}^{\Lambda}_{h\bar h}(x,\mathbf{k}_\perp) .
\label{eq:lfwf}
\end{equation}
For the radial part, we use the Gaussian wave function of the
LFQM~\cite{ChoiJi1999,ChoiJi2007},
\begin{equation}
\phi(x,\mathbf{k}_\perp)=\sqrt{\frac{\partial k_z}{\partial x}}\;
\frac{4\pi^{3/4}}{\beta^{3/2}}\,e^{-\mathbf{k}^2/2\beta^2},
\label{eq:radial}
\end{equation}
with the Jacobian $\partial k_z/\partial x=M_0/4x(1-x)$,
$\mathbf{k}=(\mathbf{k}_\perp,k_z)$, $k_z=(x-\tfrac12)M_0$ and the
invariant mass of the quark--antiquark pair is given by
\begin{equation}
M_0^2=\frac{\mathbf{k}_\perp^2+m_q^2}{x(1-x)} .
\label{eq:M0}
\end{equation}
This radial wave function is normalized as
$\int dx\,d^2\mathbf{k}_\perp\,|\phi|^2/16\pi^3=1$. The constituent masses and
the Gaussian parameters $\beta$ listed in Table~\ref{tab:par} follow from the
variational principle for the LFQM Hamiltonian (Appendix~\ref{app:var}), and the
$\psi(2S)$ has the same $m_c$ and $\beta$ as the $J/\psi$. The meson masses $M_V$ of
Table~\ref{tab:par} are used throughout, in the modified Bjorken variable, the
diffractive slope, the S-2 spin wave function, the decay constants, the leptonic
widths and the electromagnetic form factors.

\begin{table}[!t]
\caption{Model parameters used in this work (see Appendix~\ref{app:var}). Here $e_f$ is the effective quark
charge. None of these parameters is fitted to the diffractive production data.}
\label{tab:par}
\begin{ruledtabular}
\begin{tabular}{lcccc}
 & $m_q$ [GeV] & $\beta$ [GeV] & $M_V$ [GeV] & $e_f$ \\
\hline
  $\rho$ & 0.2660 & 0.3226 & 0.7753 & $1/\sqrt{2}$ \\
  $\phi$ & 0.4966 & 0.3718 & 1.0195 & $1/3$ \\
  $J/\psi$ & 1.6061 & 0.6602 & 3.0969 & $2/3$ \\
  $\psi(2S)$ & 1.6061 & 0.6602 & 3.6483 & $2/3$ \\
\end{tabular}
\end{ruledtabular}
\end{table}

\par The $J/\psi$ is described by the $1S$ wave function of Eq.~(\ref{eq:radial}). For the $\psi(2S)$, we follow Refs.~\cite{Dhiman:2019ddr,Arifi:2022pal}
and take a Gaussian multiplied by the polynomial that makes it orthogonal to the $1S$ state as
\begin{equation}
\begin{split}
\phi_{2S}(x,\mathbf{k}_\perp)=\;&\sqrt{\frac{\partial k_z}{\partial x}}\,
\frac{4\pi^{3/4}}{\beta^{3/2}}\,\mathcal{N}_{2S}\\
&\times\frac{3\beta^2-(1+2\gamma)\mathbf{k}^2}{3\beta^2}\,e^{-\gamma\mathbf{k}^2/\beta^2},
\end{split}
\label{eq:radial2S}
\end{equation}
with $\mathcal{N}_{2S}^2=96\sqrt2\,\gamma^{7/2}/(20\gamma^2-4\gamma+5)$. For $\gamma=1/2$ this is
the harmonic-oscillator $2S$ function of Ref.~\cite{Dhiman:2019ddr}, and we use $\gamma=0.7318$
from our spectrum analysis. The same Hamiltonian gives a $\psi(2S)$ mass of 3.6483~GeV
(Appendix~\ref{app:var}), which we use for $M_V$. Both charmonium states are combined with the
same spin-orbit wave functions S-1 and S-2, with $m_q=m_c$.

\par For the spin-orbit wave function, we consider two forms. The first,
S-1, follows from the Melosh--Wigner rotation of the rest-frame spin
state, which is expressed as~\cite{Melosh1974,Ji:1998zi,Puhan:2025ujg},
\begin{equation}
\mathcal{S}^{\Lambda}_{h\bar h}
 =\frac{\bar u_h(k_1)\Big[-\gamma\cdot\epsilon_\Lambda
   +\dfrac{(k_1-k_2)\cdot\epsilon_\Lambda}{M_0+2m_q}\Big]v_{\bar h}(k_2)}
   {\sqrt2\,M_0},
\label{eq:S1}
\end{equation}
where $u$ and $v$ are the light-front spinors of the quark and antiquark with four-momenta $k_1$ and $k_2$, respectively.
The polarization vector $\epsilon_\Lambda$ of the meson is evaluated with $M_0$, as
required by the Bakamjian--Thomas construction of the bound
state~\cite{BakamjianThomas1953}. Written out for the quark ($h=\uparrow,\downarrow$) and
antiquark ($\bar h=\uparrow,\downarrow$) helicities and the polarizations $\Lambda=\pm1,0$,
Eq.~(\ref{eq:S1}) becomes~\cite{Puhan:2025ujg}
\begin{align}
\mathcal{S}^{+1}&=\omega
\begin{pmatrix}
\mathbf{k}_\perp^2+AB & k^R(xM_0+m_q)\\
-k^R\big((1-x)M_0+m_q\big) & -(k^R)^2
\end{pmatrix},\nonumber\\
\mathcal{S}^{0}&=\frac{\omega}{\sqrt2}
\begin{pmatrix}
k^L C & 2\mathbf{k}_\perp^2+AB\\
2\mathbf{k}_\perp^2+AB & -k^R C
\end{pmatrix},\nonumber\\
\mathcal{S}^{-1}&=\omega
\begin{pmatrix}
-(k^L)^2 & k^L\big((1-x)M_0+m_q\big)\\
-k^L(xM_0+m_q) & \mathbf{k}_\perp^2+AB
\end{pmatrix},
\label{eq:S1mat}
\end{align}
with $k^{R,L}=k_x\pm ik_y$, $A=m_q$, $B=M_0+2m_q$, $C=(1-2x)M_0$ and
$\omega=1/\big[B\sqrt{m_q^2+\mathbf{k}_\perp^2}\big]$. This form is a unitary
rotation at every point of $(x,\mathbf{k}_\perp)$ and follow
\begin{equation}
\sum_{h\bar h}\big|\mathcal{S}^{\Lambda}_{h\bar h}\big|^2=1 ,
\label{eq:unit}
\end{equation}
so, the normalization of the full LFWF is fixed by the radial part alone.

The second form, S-2, is the ``spin-improved'' LFWF used in holographic and boosted-Gaussian
dipole calculations~\cite{ForshawSandapen2012,AhmadyEtAl2016,Gurjar:2024wpq}. We used the same
spin structure in our recent LFQM study of the $\rho$-meson distribution
functions~\cite{Puhan:2025ujg}, where its components were found to be

\begin{align}
\mathcal{S}^{+1}&=\frac{1}{2x(1-x)}
\begin{pmatrix}
m_q & x\,k^R\\
-(1-x)\,k^R & 0
\end{pmatrix},\nonumber\\
\mathcal{S}^{0}&=\frac12\left[1+\frac{m_q^2+\mathbf{k}_\perp^2}{x(1-x)M_V^2}\right]
\begin{pmatrix}
0 & 1\\ 1 & 0
\end{pmatrix},\nonumber\\
\mathcal{S}^{-1}&=\frac{1}{2x(1-x)}
\begin{pmatrix}
0 & (1-x)\,k^L\\
-x\,k^L & m_q
\end{pmatrix}.
\label{eq:S2mat}
\end{align}
The longitudinal component can also be written as
$\tfrac12\big[1+M_0^2/M_V^2\big]$, i.e.\ it contains the meson mass $M_V$
instead of $M_0$. Since Eq.~(\ref{eq:S2mat}) does not satisfy Eq.~(\ref{eq:unit}) point by
point, each polarization is normalized separately to unit $q\bar q$ probability
(Appendix~\ref{app:norm}).

\par The two structures differ most strongly in the longitudinal component.
For large $k_\perp$, the S-1 component $\mathcal{S}^0_{\uparrow\downarrow}$
approaches a constant, $\sqrt{2x(1-x)}$ ($1/\sqrt2$ at $x=1/2$), whereas the S-2 component keeps
growing, since $M_0$ becomes much larger than $M_V$ for a wide pair. The transverse components are
much closer in shape. Because S-2 is normalized only globally, however, the balance between its
longitudinal and transverse components at the dipole sizes probed by the photon also changes, and
the net effect on the cross sections has to be calculated. As shown below, S-2 gives the larger
transverse cross section and hence the smaller ratio $R$.

%======================================================================
\section{Diffractive vector-meson production}
\label{sec:diffractive}
%======================================================================

In the dipole picture, the amplitude for $\gamma^*p\to Vp$ is the convolution
of the photon and vector-meson LFWFs with the dipole--proton scattering
amplitude~\cite{KowalskiMotykaWatt2006,Kowalski:2003hm,Gurjar:2024wpq},
\begin{multline}
\mathrm{Im}\,\mathcal{A}_\Lambda(s,t;Q^2)=\sum_{h,\bar h}\int d^2\mathbf{r}_\perp\int_0^1dx\,
\Psi^{\gamma,\Lambda}_{h\bar h}(x,r_\perp;Q^2)\\
\times\Psi^{V,\Lambda*}_{h\bar h}(x,r_\perp)\,
e^{-ix\mathbf{r}_\perp\cdot\boldsymbol{\Delta}}\,
\mathcal{N}(x_m,\mathbf{r}_\perp,\boldsymbol{\Delta}),
\label{eq:ampt}
\end{multline}
where, $Q^2$ is the photon virtuality, $t=-\boldsymbol{\Delta}^2$ is the
squared momentum transfer to the proton, $\mathbf{r}_\perp$ is the transverse
size of the dipole, with complex notation $r_\perp e^{i\theta_r}$. The
dipole amplitude is evaluated at the modified Bjorken variable
\begin{equation}
x_m=x_{Bj}\left(1+\frac{M_V^2}{Q^2}\right)=\frac{Q^2+M_V^2}{W^2},
\qquad x_{Bj}=\frac{Q^2}{W^2},
\label{eq:xm}
\end{equation}
where, $W$ is the $\gamma^*p$ center-of-mass energy. At $t=0$ the integral over
impact parameter turns $\mathcal{N}$ into the dipole cross section
$\hat\sigma(x_m,r_\perp)$, and the $t$ dependence is restored through the
diffractive slope below.

\par The photon LFWFs are known from light-front QED. At lowest order, they are expressed as~\cite{LepageBrodsky1980,KowalskiMotykaWatt2006,ForshawSandapen2012,Gurjar:2024wpq}
\begin{align}
\Psi^{\gamma,0}_{h\bar h}&=\sqrt{\frac{N_c}{4\pi}}\,\delta_{h,-\bar h}\,e\,e_f\,
2x(1-x)Q\,\frac{K_0(\varepsilon r_\perp)}{2\pi},\nonumber\\
\Psi^{\gamma,\pm1}_{h\bar h}&=\pm\sqrt{\frac{N_c}{2\pi}}\,e\,e_f
\Big[ie^{\pm i\theta_r}\big(x\delta_{h\pm,\bar h\mp}\nonumber\\
&\quad-(1-x)\delta_{h\mp,\bar h\pm}\big)\partial_{r_\perp}
+m_q\delta_{h\pm,\bar h\pm}\Big]\frac{K_0(\varepsilon r_\perp)}{2\pi},
\label{eq:photon}
\end{align}
where $\varepsilon^2=x(1-x)Q^2+m_q^2$, $e^2=4\pi\alpha_{\rm em}$, $e_f$ is the
effective charge of Table~\ref{tab:par}, $N_c=3$ and $K_0$ is the modified
Bessel function. The meson LFWF is transformed to coordinate space with the
convention
$\Psi(x,\mathbf{r}_\perp)=\int d^2\mathbf{k}_\perp\,e^{-i\mathbf{k}_\perp\cdot\mathbf{r}_\perp}\Psi(x,\mathbf{k}_\perp)/(2\pi)^2$,
which is consistent with the phases of Eq.~(\ref{eq:photon}), and
after the azimuthal integration each helicity component becomes a Hankel
transform,
\begin{equation}
\begin{split}
H_n(x,r_\perp)=&\frac{1}{2\pi\sqrt{n_\Lambda}}\int_0^\infty dk_\perp\,k_\perp
J_n(k_\perp r_\perp)\\
&\times\phi(x,k_\perp)\,s(x,k_\perp),
\end{split}
\label{eq:hankel}
\end{equation}
where $s$ is the corresponding matrix element of Eq.~(\ref{eq:S1mat}) or
(\ref{eq:S2mat}) with $k^{R}$ replaced by $k_\perp$, $n$ is its azimuthal
order, and
$n_\Lambda=\sum_{h\bar h}\int dx\,d^2\mathbf{k}_\perp|\Psi^\Lambda_{h\bar h}|^2/(2\pi)^2$
ensures $\sum_{h\bar h}\int dx\,d^2\mathbf{r}_\perp|\Psi^\Lambda_{h\bar h}(x,\mathbf{r}_\perp)|^2=1$.
We use $H_L$ for $\mathcal{S}^0_{\uparrow\downarrow}$ ($n=0$), $H_{1a}$ and
$H_{1b}$ for $\mathcal{S}^{+1}_{\uparrow\downarrow}$ and
$-\mathcal{S}^{+1}_{\downarrow\uparrow}$ ($n=1$), and $H_{0c}$ for
$\mathcal{S}^{+1}_{\uparrow\uparrow}$ ($n=0$). Summed over helicities, the
photon--meson overlaps are found to be
\begin{align}
\big(\Psi_V^*\Psi_\gamma\big)_L&=2\sqrt{\frac{N_c}{4\pi}}\,e\,e_f\,
  \frac{2x(1-x)Q}{2\pi}\,K_0(\varepsilon r_\perp)\,H_L(x,r_\perp),
  \nonumber\\
\big(\Psi_V^*\Psi_\gamma\big)_T&=\sqrt{\frac{N_c}{2\pi}}\,\frac{e\,e_f}{2\pi}
  \Big\{\varepsilon K_1(\varepsilon r_\perp)\big[xH_{1a}+(1-x)H_{1b}\big]
  \nonumber\\
  &\qquad\qquad+m_q K_0(\varepsilon r_\perp)H_{0c}\Big\},
\label{eq:overlap}
\end{align}
where $\partial_{r_\perp}K_0(\varepsilon r_\perp)=-\varepsilon K_1(\varepsilon r_\perp)$
has been used. Only the helicity components that the photon can reach
contribute. The $h=\bar h$ components of $\mathcal{S}^0$ and the
$(\downarrow\downarrow)$ component of $\mathcal{S}^{+1}$ in S-1 therefore drop out of
Eq.~(\ref{eq:overlap}), although they are kept in the normalization.
Eq.~(\ref{eq:overlap}) therefore holds for both spin structures, but the
functions $H$, and hence the explicit overlaps, are different. For S-2, all
four $H$ follow from a single scalar function $\Psi(x,r_\perp)$, the transform
of Eq.~(\ref{eq:hankel}) with $n=0$ and $s=1$ (normalized separately for
$\Lambda=0$ and $\pm1$), and Eq.~(\ref{eq:overlap}) reduces to the familiar
form as~\cite{ForshawSandapen2012,Gurjar:2024wpq}
\begin{align}
\big(\Psi_V^*\Psi_\gamma\big)^{\text{S-2}}_L&=\sqrt{\frac{N_c}{4\pi}}\,e\,e_f\,
  \frac{2x(1-x)Q}{2\pi}K_0(\varepsilon r_\perp)\nonumber\\
&\quad\times\Big[1+\frac{m_q^2-\nabla^2_{r_\perp}}{x(1-x)M_V^2}\Big]\Psi,\nonumber\\
\big(\Psi_V^*\Psi_\gamma\big)^{\text{S-2}}_T&=\sqrt{\frac{N_c}{2\pi}}\,\frac{e\,e_f}{2\pi}\,
  \frac{1}{2x(1-x)}\Big\{m_q^2K_0\Psi\nonumber\\
&\quad-\big[x^2+(1-x)^2\big]\varepsilon K_1\,\partial_{r_\perp}\Psi\Big\},
\label{eq:overlapS2}
\end{align}
with $K_{0,1}\equiv K_{0,1}(\varepsilon r_\perp)$. For S-1, the components
depend on $k_\perp$ also through $M_0$ and $\omega$, so no such single
function exists and the overlaps remain integrals over $k\equiv k_\perp$,
\begin{align}
\big(\Psi_V^*\Psi_\gamma\big)^{\text{S-1}}_L&=\sqrt{\frac{N_c}{4\pi}}\,e\,e_f\,
  \frac{2x(1-x)Q}{2\pi}K_0\nonumber\\
&\quad\times\int_0^\infty\!\frac{dk\,k}{\pi\sqrt{n_0}}\,J_0(kr_\perp)\,
  \phi\,\frac{\omega(2k^2+AB)}{\sqrt2},\nonumber\\
\big(\Psi_V^*\Psi_\gamma\big)^{\text{S-1}}_T&=\sqrt{\frac{N_c}{2\pi}}\,\frac{e\,e_f}{2\pi}
  \int_0^\infty\!\frac{dk\,k}{2\pi\sqrt{n_1}}\,\phi\,\omega\nonumber\\
&\quad\times\Big\{\varepsilon K_1J_1(kr_\perp)\,k\big[(x^2+(1-x)^2)M_0\nonumber\\
&\qquad+m_q\big]\nonumber\\
&\qquad+m_qK_0J_0(kr_\perp)(k^2+AB)\Big\}.
\label{eq:overlapS1}
\end{align}
The difference between Eqs.~(\ref{eq:overlapS2}) and (\ref{eq:overlapS1}) is
the only difference between the two calculations. The forward amplitude is
then
\begin{equation}
\begin{split}
\mathrm{Im}\,\mathcal{A}_\Lambda(Q^2,x_m)
 =&\int_0^1\!dx\int d^2\mathbf{r}_\perp\,
 \big(\Psi_V^*\Psi_\gamma\big)_\Lambda(x,r_\perp)\\
 &\times\hat\sigma(x_m,r_\perp).
\end{split}
\label{eq:amp}
\end{equation}

\par For the dipole cross section, we use the color-glass-condensate (CGC)
parametrization of Ref.~\cite{IancuItakuraMunier2004},
$\hat\sigma(x_m,r_\perp)=\sigma_0\,\mathcal{N}(x_m,r_\perp Q_s)$ with
\begin{equation}
\mathcal{N}=
\begin{cases}
\mathcal{N}_0\left(\dfrac{r_\perp Q_s}{2}\right)^{2\left[\gamma_s+\frac{\ln(2/r_\perp Q_s)}{\kappa_0\lambda\ln(1/x_m)}\right]}, & r_\perp Q_s\le2,\\[2ex]
1-\exp\!\big[-\mathcal{A}\ln^2(\mathcal{B}\,r_\perp Q_s)\big], & r_\perp Q_s>2,
\end{cases}
\label{eq:cgc}
\end{equation}
and saturation scale $Q_s=(x_0/x_m)^{\lambda/2}$~GeV. The coefficients
\begin{equation}
\mathcal{A}=-\frac{\mathcal{N}_0^2\gamma_s^2}{(1-\mathcal{N}_0)^2\ln(1-\mathcal{N}_0)},\qquad
\mathcal{B}=\frac12(1-\mathcal{N}_0)^{-\frac{1-\mathcal{N}_0}{\mathcal{N}_0\gamma_s}}
\label{eq:AB}
\end{equation}
make $\mathcal{N}$ and its derivative continuous at $r_\perp Q_s=2$. The
parameters $\sigma_0$, $\gamma_s$, $\lambda$ and $x_0$ are fitted to inclusive
deep-inelastic data. Earlier fits~\cite{IancuItakuraMunier2004,Soyez:2007kg,Rezaeian2013,AhmadyEtAl2016}
used current or effective light-quark masses, $m_{u,d}\le0.14$~GeV, in the photon wave
function. Because the same photon wave function, Eq.~(\ref{eq:photon}), enters the vector-meson
amplitude with the LFQM masses of Table~\ref{tab:par}, we determine the parameters by fitting the
structure-function data with $m_{u,d}=0.266$, $m_s=0.497$ and $m_c=1.606$~GeV. These are the
combined H1 and ZEUS inclusive~\cite{H1ZEUS:2015HERA} and charm~\cite{H1:2018flt} reduced cross
sections and the proton structure function $F_2$ measured by E665~\cite{E665:1996mob} and
NMC~\cite{NewMuon:1996fwh}, all with $x_{Bj}\le0.01$ and $0.045\le Q^2\le45$~GeV$^2$.
With $\mathcal{N}_0=\Nzero$ and $\kappa_0=\kappaZero$ (the leading-order BFKL value) held fixed,
we obtain $\sigma_0=\sigmaZero$~mb, $\gamma_s=\gammaS$, $\lambda=\lambdaCGC$ and $x_0=\xZero$,
with $\chi^2/N=\cgcChiN$. The fit is described in Appendix~\ref{app:cgc}.

\par For a given photon polarization, the differential and integrated cross sections
are then
\begin{gather}
\frac{d\sigma_\Lambda}{d|t|}=\frac{\big[\mathrm{Im}\,\mathcal{A}_\Lambda\big]^2}{16\pi}
(1+\beta_\Lambda^2)\,e^{-B_D|t|},\nonumber\\
\sigma_\Lambda=\frac{\big[\mathrm{Im}\,\mathcal{A}_\Lambda\big]^2(1+\beta_\Lambda^2)}{16\pi B_D},
\label{eq:sigma}
\end{gather}
where, the real part of the amplitude is included through~\cite{KowalskiMotykaWatt2006}
\begin{equation}
\beta_\Lambda=\tan\left(\frac{\pi\alpha_\Lambda}{2}\right),\qquad
\alpha_\Lambda=\frac{\partial\ln|\mathrm{Im}\,\mathcal{A}_\Lambda|}{\partial\ln(1/x_m)} ,
\label{eq:beta}
\end{equation}
and the diffractive slope is~\cite{AhmadyEtAl2016,Gurjar:2024wpq}
\begin{equation}
B_D=N\left[14\left(\frac{1~\mathrm{GeV}^2}{Q^2+M_V^2}\right)^{0.2}+1\right],
\quad N=0.55~\mathrm{GeV}^{-2}.
\label{eq:BD}
\end{equation}
No skewedness correction is applied. The measured cross section is
\begin{equation}
\sigma=\sigma_T+\epsilon\,\sigma_L ,
\label{eq:sigtot}
\end{equation}
where, $\epsilon$ is the longitudinal polarization of the virtual photon. It is close to one at
HERA ($\epsilon\simeq\omegaEps$) but ranges from 0.5 to 0.84 in the fixed-target experiments, so
the calculation is always evaluated at the $W$ and $\epsilon$ of the data set it is compared
with.

\par
\begin{figure*}[t]
\centering
\includegraphics[width=\textwidth]{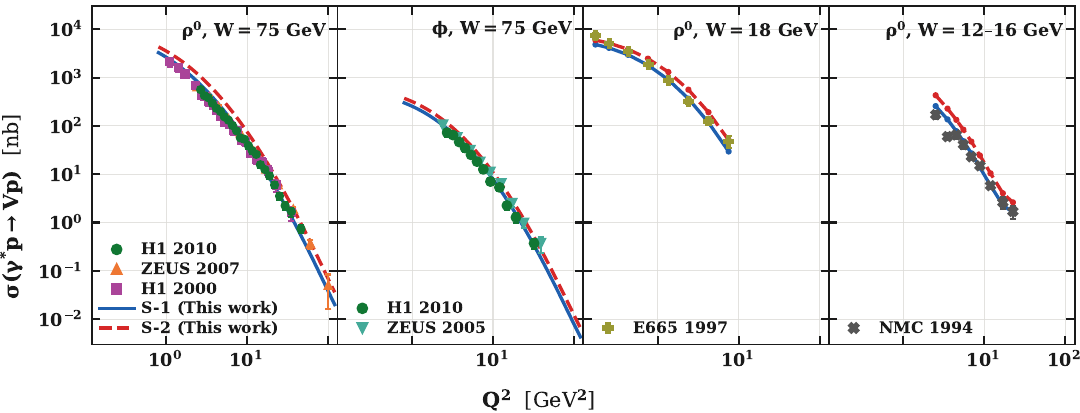}
\caption{(Color online) Total cross section $\sigma(\gamma^*p\to Vp)$ as a
function of $Q^2$. The panels show, from left to right, the $\rho^0$ and $\phi$ at $W=75$~GeV
compared with H1~\cite{H1rho2010,H1rho2000} and
ZEUS~\cite{ZEUSrho2007,ZEUSphi2005}, and the $\rho^0$ at fixed-target energies
compared with E665~\cite{E665rho1997} and NMC~\cite{NMCrho1994}. The solid
(blue) and dashed (red) lines are the S-1 and S-2 results. The calculation is
evaluated at the $W$ and $\epsilon$ of each data set. Since $W$ and $\epsilon$
change from point to point for NMC, the curves in that panel are joined
point-by-point results.}
\label{fig:sigQ2}
\end{figure*}

\begin{figure*}[t]
\centering
\includegraphics[width=\textwidth]{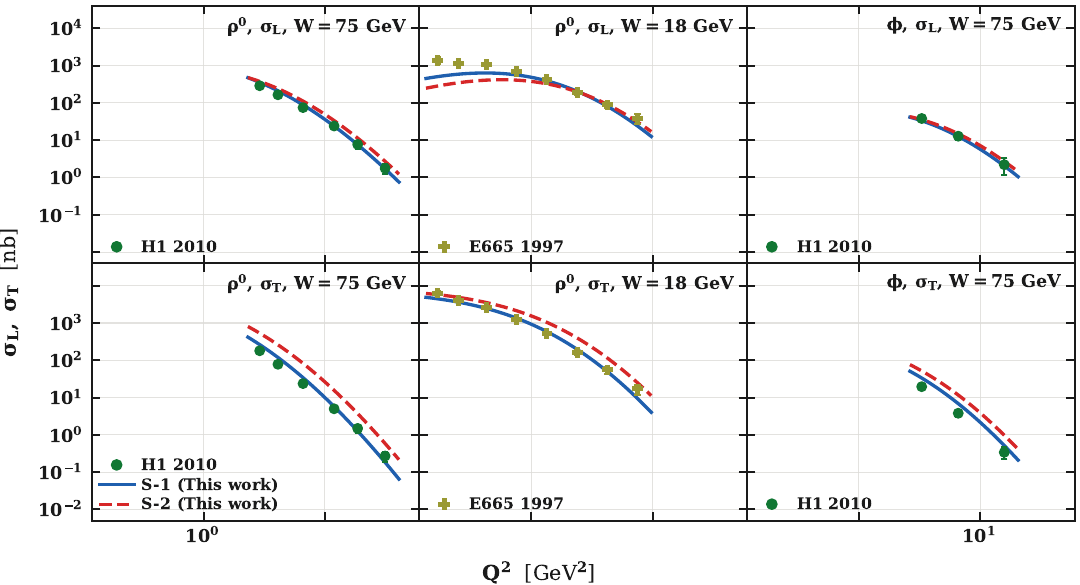}
\caption{(Color online) Longitudinal (upper row) and transverse (lower row)
cross sections of the $\rho^0$ at $W=75$~GeV (H1~\cite{H1rho2010}) and at
$W=18$~GeV (E665~\cite{E665rho1997}), and of the $\phi$ at $W=75$~GeV
(H1~\cite{H1rho2010}). Lines as in Fig.~\ref{fig:sigQ2}.}
\label{fig:sigLT}
\end{figure*}

\begin{figure*}[t]
\centering
\includegraphics[width=\textwidth]{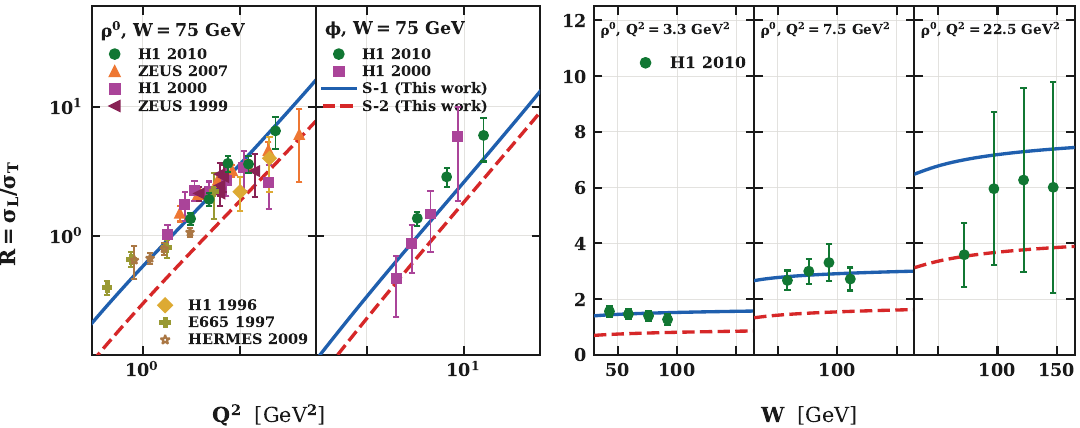}
\caption{(Color online) Ratio $R=\sigma_L/\sigma_T$ as a function of
$Q^2$ for the $\rho^0$ and the $\phi$ (left), compared with
H1~\cite{H1rho2010,H1rho2000,H1rho1996},
ZEUS~\cite{ZEUSrho2007,ZEUSrho1999}, E665~\cite{E665rho1997} and
HERMES~\cite{HERMESrho2009}, and $R$ for the $\rho^0$ as a function of $W$
at three values of $Q^2$ (right), compared with H1~\cite{H1rho2010}. Lines as in
Fig.~\ref{fig:sigQ2}.}
\label{fig:R}
\end{figure*}

\begin{figure*}[t]
\centering
\includegraphics[width=\textwidth]{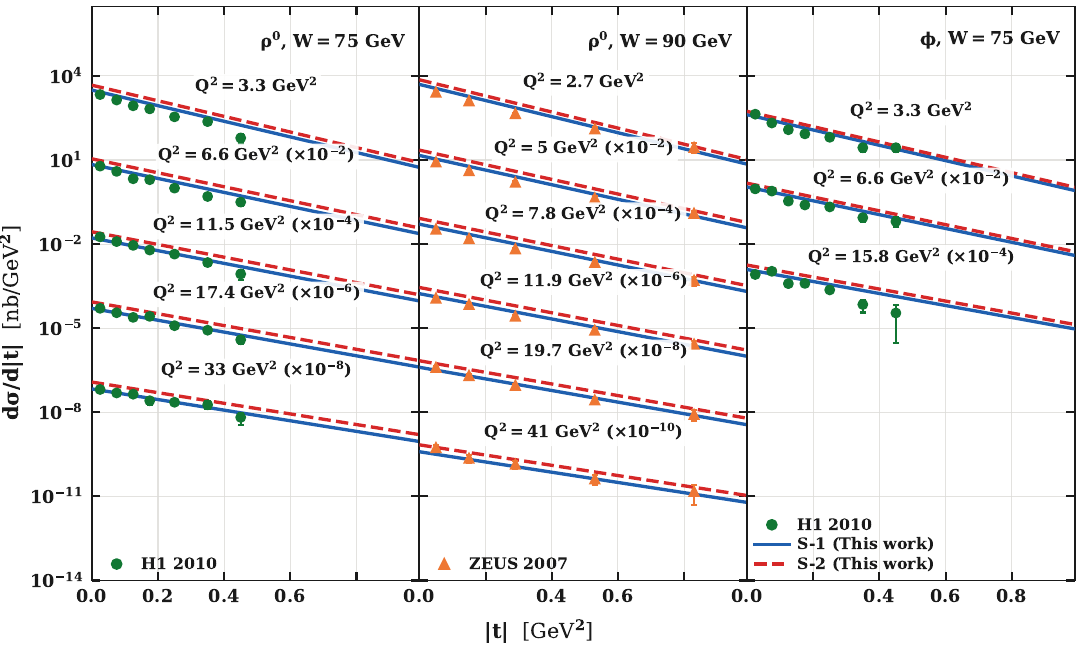}
\caption{(Color online) Differential cross section $d\sigma/d|t|$ in the
measured $Q^2$ bins for the $\rho^0$ at $W=75$~GeV (H1~\cite{H1rho2010}) and
$W=90$~GeV (ZEUS~\cite{ZEUSrho2007}), and for the $\phi$ at $W=75$~GeV
(H1~\cite{H1rho2010}). Successive bins are scaled by the factors indicated.
Lines as in Fig.~\ref{fig:sigQ2}.}
\label{fig:dsdt}
\end{figure*}

\begin{figure*}[t]
\centering
\includegraphics[width=\textwidth]{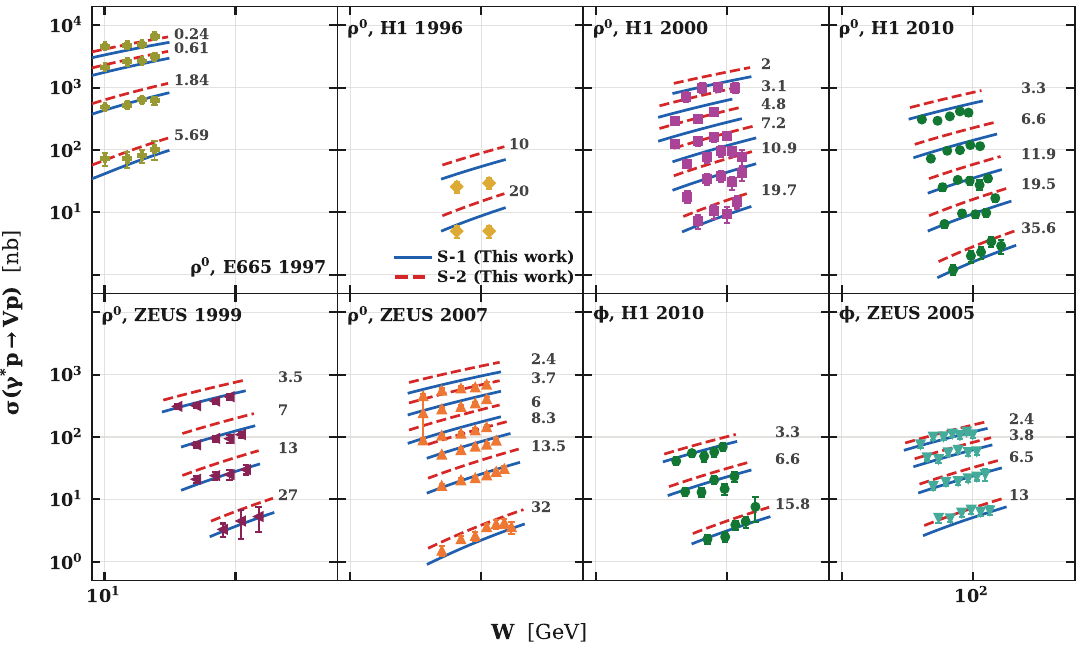}
\caption{(Color online) Total cross section as a function of $W$ at fixed
$Q^2$ for all the $\rho^0$ and $\phi$ data sets of E665~\cite{E665rho1997},
H1~\cite{H1rho1996,H1rho2000,H1rho2010} and
ZEUS~\cite{ZEUSrho1999,ZEUSrho2007,ZEUSphi2005}. The number next to each band
is its $Q^2$ in GeV$^2$, and each curve is drawn over the $W$ range of its
data. E665 quotes $\sigma_T+\sigma_L$, and the curves in that panel are
calculated accordingly. Lines as in Fig.~\ref{fig:sigQ2}.}
\label{fig:sigW}
\end{figure*}

\begin{figure}[!t]
\centering
\includegraphics[width=\columnwidth]{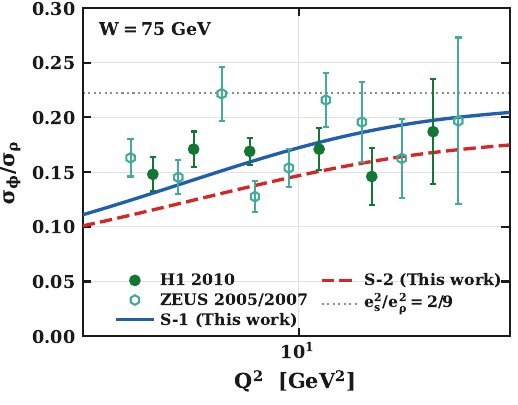}
\caption{(Color online) Ratio $\sigma_\phi/\sigma_\rho$ as a function of $Q^2$
at $W=75$~GeV, compared with the ratio measured by H1~\cite{H1rho2010} (filled
circles) and the ratio formed from the separate ZEUS
measurements~\cite{ZEUSrho2007,ZEUSphi2005} (open symbols). The dotted line is
$e_s^2/e_\rho^2=2/9$.}
\label{fig:ratio}
\end{figure}

\begin{figure*}[t]
\centering
\includegraphics[width=0.67\textwidth]{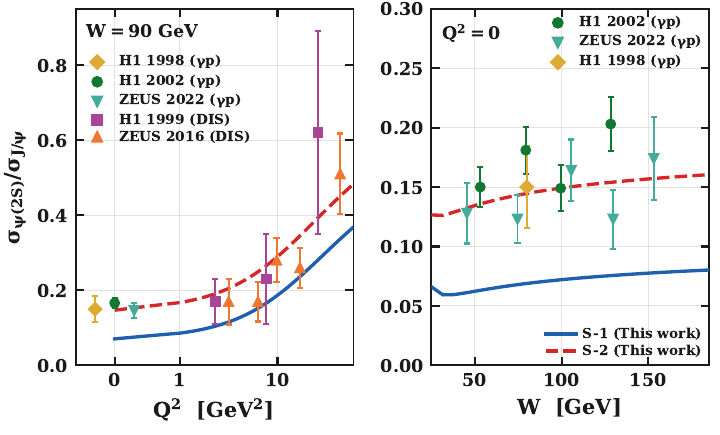}
\caption{(Color online) Ratio $\sigma_{\psi(2S)}/\sigma_{J/\psi}$ as a function of $Q^2$ at
$W=90$~GeV (left) and of $W$ at $Q^2=0$ (right), compared with H1~\cite{H1:1998psi,H1:1999psi,H1:2002psi} and
ZEUS~\cite{ZEUS:2016psi,ZEUS:2022psi}. Lines as in Fig.~\ref{fig:sigQ2}.}
\label{fig:psi}
\end{figure*}

 We compare the results with the data of H1~\cite{H1rho2010,H1rho2000,H1rho1996},
ZEUS~\cite{ZEUSrho2007,ZEUSrho1999,ZEUSphi2005}, E665~\cite{E665rho1997}, NMC~\cite{NMCrho1994}
and HERMES~\cite{HERMESrho2009}. For the $\rho^0$ and $\phi$, \nPoints\ data points from
\nDatasets\ data sets are included. They cover the total cross sections as functions of $Q^2$
and $W$, the separated longitudinal and transverse cross sections, the ratio $R$ as a function of $Q^2$ and
$W$, the $t$ distributions in several $Q^2$ bins, and the ratio $\sigma_\phi/\sigma_\rho$. For
the $J/\psi$ and $\psi(2S)$, we use the ratio $\sigma_{\psi(2S)}/\sigma_{J/\psi}$ measured by
H1~\cite{H1:1998psi,H1:1999psi,H1:2002psi} and ZEUS~\cite{ZEUS:2016psi,ZEUS:2022psi}. All the
data points are shown in Figs.~\ref{fig:sigQ2}--\ref{fig:psi}. Where an experiment quotes the
spin-density matrix element $r^{04}_{00}$ instead of $R$, we convert it to $R$ assuming
$s$-channel helicity conservation (SCHC), with the $\epsilon$ of that experiment.

\par In Fig.~\ref{fig:sigQ2}, we show the total cross section as a function of $Q^2$ for the
$\rho^0$ and $\phi$ at HERA energies, together with the fixed-target $\rho^0$ data of E665 and
NMC. The calculations are not normalized to the data. For both mesons the $Q^2$ dependence is well
described over more than two decades in $Q^2$ and nearly four orders of magnitude in the cross
section, but the two spin structures give different normalizations. The S-1 results lie
close to the HERA data, whereas S-2 lies above them. The ratio $N=\sigma_{\rm exp}/\sigma_{\rm th}$
averaged over the H1 2010 $\rho$ points is $\normrhoSOne\pm\normSpreadrhoSOne$ for S-1 and
$\normrhoSTwo\pm\normSpreadrhoSTwo$ for S-2, and the other HERA data sets give 0.84--1.23 and
0.54--0.88, respectively. For the $\rho^0$, $N$ grows with $Q^2$, from about 0.8 in the lowest
third of the measured range to 1.1--1.3 in the highest third for S-1 (from 0.5 to 0.65--0.75 for
S-2), so the calculated cross section falls somewhat faster with $Q^2$ than the data. For the H1
$\phi$ data, $N$ is found to be constant. The difference between S-1 and S-2 is nevertheless an overall
factor. Although $\sigma_0$ is now fitted with the same quark masses, the normalization still
depends on the skewedness correction, which is not included and would raise both calculations by
the same factor, and on the diffractive slope $B_D$. The normalization therefore supports S-1, but
we do not use it on its own to discriminate between the two wave functions. Among the
fixed-target data, the NMC points at $W=12$--16~GeV lie close to S-1 ($N=0.44$--1.14), while S-2
lies above them by a factor of 1.5--3.8. The E665 points at $W=18$~GeV fall between the two
calculations, with $N=0.9$--1.6 for S-1 and 0.6--1.25 for S-2, and $N$ grows towards the lowest
$Q^2$. At such low energies and virtualities, contributions beyond two-gluon exchange, such as
secondary-Reggeon exchange~\cite{Donnachie:1992ny}, and large nonperturbative dipoles are
expected to matter.

\par In Fig.~\ref{fig:sigLT}, we demonstrate the longitudinal and transverse cross sections. For the H1
data at $W=75$~GeV, the two spin structures behave differently in the two polarizations. They are
close to each other for $\sigma_L$, whereas for $\sigma_T$ S-2 lies clearly above S-1. This is
what makes the ratio $R$ sensitive to the spin structure. The E665 measurements at $W=18$~GeV
extend down to $Q^2=0.17$~GeV$^2$. At the lowest $Q^2$, $\sigma_T$ is described by S-2
($N=1.1$) and underestimated by S-1 by a factor of 1.4, while $\sigma_L$ is underestimated by both
calculations, by factors of 2.8 (S-1) and 4.9 (S-2), in line with what we found for the total
cross section.

\par In the ratio $R=\sigma_L/\sigma_T$ the normalization $\sigma_0$, the skewedness factor
and the slope $B_D$ cancel, since they multiply both polarizations in the same way, and $R$
depends only on the relative weight of the longitudinal and transverse components of the meson
wave function. Because $R$ is dimensionless, measurements made at different $W$ and $\epsilon$ can
be compared with a single curve. The left panels of Fig.~\ref{fig:R}, show $R$ as a function of
$Q^2$ for both mesons together with all the available measurements. On the logarithmic scale the
two spin structures give nearly parallel straight lines, i.e., they differ by an almost constant
factor over the whole $Q^2$ range. At $Q^2=6.6$~GeV$^2$, $R_{\text{S-1}}/R_{\text{S-2}}$ is
\discrho\ for the $\rho$ and \discphi\ for the $\phi$. Between $Q^2=3.3$ and
19.7~GeV$^2$, the relative difference $(R_{\text{S-1}}-R_{\text{S-2}})/R_{\text{S-2}}$ grows
slowly from 91\% to 97\% for the $\rho$ and stays at 43--45\% for the $\phi$.

\par The data lie clearly closer to S-1. To quantify this, we calculate $\chi^2$ for the 34
HERA points without fitting any parameter (Table~\ref{tab:chi2}). S-1 gives
$\chi^2=\chiTwoSOneRaw$, i.e.\ \chiTwoSOne\ per point, and S-2 gives $\chi^2=\chiTwoSTwoRaw$,
i.e.\ \chiTwoSTwo\ per point. Most of the individual data sets also prefer S-1. The exceptions
are the H1 2000 $\phi$ data, which cannot distinguish between the two, and the two H1 1996
points, which favor S-2. The E665 and HERMES points are left out of the $\chi^2$, because they
were measured at $W\approx5$--20~GeV, where contributions other than two-gluon exchange are not
negligible. For HERMES, moreover, $x_m$ lies well above $10^{-2}$, outside the range of the dipole
fit.

\par The right panels of Fig.~\ref{fig:R} show $R$ as a function of $W$ at three values of
$Q^2$. Both calculations are nearly independent of $W$, because $W$ enters only through $x_m$ in
the dipole cross section, which largely cancels in the ratio. The H1 data show no significant $W$
dependence either, which supports the factorization of the amplitude into the meson wave function
and the dipole cross section. In the two lower $Q^2$ bins, the data are close to S-1 at all $W$,
while at $Q^2=22.5$~GeV$^2$ they lie between the two curves, with large uncertainties.

\begin{table}[!t]
\caption{$\chi^2/N$ of the measured $R$ for the two spin wave functions (HERA data only) presented in Fig. \ref{fig:R}.}
\label{tab:chi2}
\begin{ruledtabular}
% ---- tabbody_chi2 (inlined) ----
\begin{tabular}{llccc}
meson & dataset & points & $\chi^2/N$ (S-1) &
 $\chi^2/N$ (S-2) \\
\hline
$\rho^{0}$ & H1 2010 & 5 & 0.84 & 11.80 \\
  $\rho^{0}$ & ZEUS 2007 & 7 & 2.89 & 16.31 \\
  $\rho^{0}$ & H1 2000 & 7 & 2.80 & 5.87 \\
  $\rho^{0}$ & ZEUS 1999 & 6 & 1.01 & 7.41 \\
  $\rho^{0}$ & H1 1996 & 2 & 3.41 & 0.17 \\
  $\phi$ & H1 2010 & 3 & 3.55 & 9.95 \\
  $\phi$ & H1 2000 & 4 & 0.34 & 0.58 \\
\hline
 & all & 34 & \chiTwoSOne &
 \chiTwoSTwo \\
\end{tabular}
% ---- end tabbody_chi2 ----
\end{ruledtabular}
\end{table}

\par Figure~\ref{fig:dsdt} shows $d\sigma/d|t|$ in all the measured $Q^2$ bins. The slope is
not fitted but taken from Eq.~(\ref{eq:BD}) at each $Q^2$. The exponential fall-off of the data is
well described in all the bins. Since the $t$ dependence is governed entirely by $B_D$ and not by
the meson wave function, the two spin structures differ only in normalization, and these data
test the overall framework without distinguishing S-1 from S-2.

\par Figure~\ref{fig:sigW} shows the energy dependence of the cross section at fixed $Q^2$ for
all the data sets, from the E665 measurement at $W\approx10$~GeV up to the HERA data at
$W\approx170$~GeV. The rise with $W$ is generated by the saturation scale through $x_m$ and is well
described at all $Q^2$. As in Fig.~\ref{fig:sigQ2}, S-1 and S-2 differ mainly by an overall factor,
which shows that the rise with $W$ is a property of the dipole cross section rather than of the
meson wave function.

\par The ratio of the $\phi$ and $\rho^0$ cross sections is shown in Fig.~\ref{fig:ratio}. The
dipole normalization cancels here as well, but unlike $R$ this ratio changes the quark mass and
charge at a fixed radial structure, and so tests the quark-mass dependence of the spin structure.
Both calculations reproduce the slow rise of the measured ratio from about 0.10 to 0.20 and stay
below the value $2/9$ expected from the quark charges alone. Since both describe the data within
their uncertainties, this ratio does not discriminate between S-1 and S-2.

\par For the $J/\psi$ and $\psi(2S)$, we use the same photon wave function, dipole cross section,
real-part correction and slope as for the light mesons. The charm quark enters the dipole fit
with $m_c=1.606$~GeV, but the calculated $J/\psi$ photoproduction cross section at $W=90$~GeV,
38~nb (S-1) and 43~nb (S-2), is below the H1 value of $73\pm7$~nb~\cite{H1:2005jpsi}. We
therefore consider only the ratio $\sigma_{\psi(2S)}/\sigma_{J/\psi}$, in which the normalization
cancels. The ratio is sensitive to the node of the $\psi(2S)$ wave function, since dipoles smaller
and larger than the node contribute with opposite signs~\cite{Nemchik:1997xb,Hoyer:1999xe,Suzuki:2000az}.

\par Figure~\ref{fig:psi} shows the ratio as a function of $Q^2$ and $W$ together with the
H1~\cite{H1:1998psi,H1:1999psi,H1:2002psi} and ZEUS~\cite{ZEUS:2016psi,ZEUS:2022psi} data. S-2 describes the
data well in both variables. At $Q^2=0$ and $W=90$~GeV it gives 0.15, in agreement with the
measured 0.150 (H1 1998~\cite{H1:1998psi}), 0.166 (H1 2002~\cite{H1:2002psi}) and 0.146 (ZEUS
2022~\cite{ZEUS:2022psi}), whereas S-1 gives only 0.07. The ratio rises with $Q^2$, because
the dipoles become smaller and the cancellation around the node weakens. S-2 gives 0.29 at
$Q^2=10$~GeV$^2$ and 0.41 at 30~GeV$^2$, consistent with the DIS data, while S-1 lies below
them. The S-1 spin structures give a weak $W$ dependence, while the $S-2$ matches with the data .

\par The ratio also depends on the shape of the radially excited wave function, and a different
$2S$ function or $1S$--$2S$ mixing~\cite{Arifi:2022pal} can change it considerably. Unlike $R$ for
the light mesons, the charmonium ratio therefore tests the spin and the radial structure of the
$\psi(2S)$ together, as other dipole studies have also found
\cite{Cepila:2019skb,Krelina:2019egg,Henkels:2020kju,Jamal:2026heavy}.

%======================================================================
\section{Decay constants and electromagnetic form factors}
\label{sec:emff}
%======================================================================
\par The decay constant of a vector meson is defined by
$\langle0|\bar q\gamma^\mu q|V(P,\Lambda)\rangle=f_VM_V\epsilon^\mu_\Lambda$ \cite{Arifi:2022pal}.
Taking the plus component and $\Lambda=0$, it follows from the same LFWF as
\begin{equation}
f_V=4\sqrt{N_c}\int\frac{dx\,d^2\mathbf{k}_\perp}{16\pi^3}\,
\phi(x,\mathbf{k}_\perp)\,\mathcal{S}^{0}_{\uparrow\downarrow}(x,\mathbf{k}_\perp),
\label{eq:fv}
\end{equation}
which for the two spin structures is found to be
\begin{align}
f_V^{\text{S-1}}&=2\sqrt6\int\frac{dx\,d^2\mathbf{k}_\perp}{16\pi^3}\,
\phi(x,\mathbf{k}_\perp)\,\frac{m_q+2\mathbf{k}_\perp^2/B}{\sqrt{m_q^2+\mathbf{k}_\perp^2}},
\label{eq:fvS1}\\
f_V^{\text{S-2}}&=\frac{2\sqrt3}{\sqrt{\mathcal{N}_L}}\int\frac{dx\,d^2\mathbf{k}_\perp}{16\pi^3}\,
\phi(x,\mathbf{k}_\perp)\left[1+\frac{M_0^2}{M_V^2}\right],
\label{eq:fvS2}
\end{align}
where $B=M_0+2m_q$ and $\mathcal{N}_L$ is the normalization of the longitudinal S-2
state, given in Eq.~(\ref{eq:NLNT}) and Table~\ref{tab:norm} of Appendix~\ref{app:norm}. Eq.~(\ref{eq:fvS1}) is the familiar LFQM
result~\cite{ChoiJi2007}. The leptonic width is
$\Gamma(V\to e^+e^-)=4\pi\alpha_{\rm em}^2C_V^2f_V^2/(3M_V)$, with
$C_\rho^2=1/2$, $C_\phi^2=1/9$ and $C_{J/\psi}^2=C_{\psi(2S)}^2=4/9$.

\par The decay constants test the spin structure independently, since they do not involve the
dipole cross section. Our results are collected in Table~\ref{tab:static}, together with the
masses, leptonic widths and electromagnetic static properties. S-1 gives $f_\rho\approx217$~MeV and
$f_\phi\approx241$~MeV, within 2\% and 6\% of the values 221 and 229~MeV extracted from the
leptonic widths \cite{PDG2024}, whereas S-2 gives about 273 and 290~MeV, i.e.\ 23\% and 27\% too large. The
decay constants therefore point to the same conclusion as $R$, without any dependence on the
dipole model. The corresponding leptonic widths are 6.76 and 1.41~keV for S-1 and 10.70 and
2.05~keV for S-2, compared with the measured 7.04 and 1.27~keV~\cite{PDG2024}. The meson masses
and radial wave functions are the same for both spin structures.

\par Charmonium behaves differently. S-2 gives $f_{J/\psi}=416$~MeV and $\Gamma_{ee}=5.54$~keV, in
agreement with the measured 416~MeV and 5.53~keV~\cite{PDG2024}, while S-1 gives 379~MeV and
4.60~keV, i.e.\ 9\% below the measured decay constant. For the $\psi(2S)$, S-1 and S-2 give 246
and 301~MeV, compared with the measured 294~MeV~\cite{PDG2024}. The ratio
$f_{\psi(2S)}/f_{J/\psi}$ is 0.65 with S-1 and 0.72 with S-2, against 0.71 from experiment, which
is the same trend as in $\sigma_{\psi(2S)}/\sigma_{J/\psi}$ (Fig.~\ref{fig:psi}).

\begin{table*}[!t]
\caption{Static properties of the four mesons with S-1 and S-2, compared with
experiment~\cite{PDG2024} and LQCD. Here $\mu_V$ is in units of $e/2M_V$ and
$Q_V=G_Q(0)/M_V^2$. The lattice $\rho^+$ results are at $m_\pi\approx700$~MeV
(HadSpec~\cite{Shultz:2015pfa}) and 406~MeV (QCDSF~\cite{QCDSF:2008tjq}).}
\label{tab:static}
\small
\begin{ruledtabular}
\begin{tabular}{llccccccc}
 & & $M_V$ [MeV] & $f_V$ [MeV] & $\Gamma_{ee}$ [keV] & $r_C$ [fm] & $\mu_V$ [$e/2M_V$] & $G_Q(0)$ & $Q_V$ [fm$^2$] \\
\hline
$\rho$ & S-1 & 775.3 & 216.7 & 6.76 & 0.67 & 2.21 & $-0.25$ & $-0.016$ \\
 & S-2 & 775.3 & 272.7 & 10.70 & 0.78 & 2.23 & $-0.71$ & $-0.046$ \\
 & Expt.~\cite{PDG2024} & 775.3 & 221.2(1.0) & 7.04(6) & & & & \\
 & LQCD (HadSpec)~\cite{Shultz:2015pfa} & & & & 0.55(5) & 2.17(10) & $-0.54(10)$ & $-0.020(4)$ \\
 & LQCD (QCDSF)~\cite{QCDSF:2008tjq} & & & & 0.70(3) & 1.7(3) & --- & $-0.015(4)$ \\
\hline
$\phi$ & S-1 & 1019.5 & 241.1 & 1.41 & 0.48 & 2.12 & $-0.42$ & $-0.016$ \\
 & S-2 & 1019.5 & 290.1 & 2.05 & 0.50 & 2.13 & $-0.75$ & $-0.028$ \\
 & Expt.~\cite{PDG2024} & 1019.5 & 228.6(3.6) & 1.27(4) & & & & \\
\hline
$J/\psi$ & S-1 & 3096.9 & 379.1 & 4.60 & 0.22 & 2.06 & $-0.26$ & $-0.0011$ \\
 & S-2 & 3096.9 & 416.2 & 5.54 & 0.21 & 2.06 & $-0.64$ & $-0.0026$ \\
 & Expt.~\cite{PDG2024} & 3096.9 & 415.6(3.8) & 5.53(10) & & & & \\
 & LQCD ($N_f=0$)~\cite{Dudek:2006ej} & & & & 0.257(4) & 2.10(3) & $-0.23(2)$ & $-0.0009(1)$ \\
 & LQCD ($N_f=2+1$)~\cite{Delaney:2023fsx} & & & & 0.249(4) & 2.287(18) & $-0.172(15)$ & $-0.0007(1)$ \\
\hline
$\psi(2S)$ & S-1 & 3648.3 & 245.5 & 1.64 & 0.39 & 2.07 & $-0.01$ & $-0.00003$ \\
 & S-2 & 3648.3 & 301.1 & 2.46 & 0.37 & 2.09 & $+0.21$ & $+0.0006$ \\
 & Expt.~\cite{PDG2024} & 3686.1 & 294.3(2.5) & 2.33(4) & & & & \\
\end{tabular}
\end{ruledtabular}
\end{table*}

\par The electromagnetic current of a spin-1 particle is described by three
form factors $G_{1,2,3}$($Q^2$) and can be calculated from the correlation function between the initial and final meson with vector current $J^\mu$ as~\cite{Arnold:1979cg}
\begin{multline}
\langle V(P',\Lambda')|J^\mu|V(P,\Lambda)\rangle
=-\Big\{G_1(Q^2)\,(\epsilon'^*\!\cdot\epsilon)\,(P+P')^\mu\\
+G_2(Q^2)\big[\epsilon^\mu(\epsilon'^*\!\cdot q)-\epsilon'^{*\mu}(\epsilon\cdot q)\big]\\
-G_3(Q^2)\,\frac{(\epsilon'^*\!\cdot q)(\epsilon\cdot q)}{2M_V^2}(P+P')^\mu\Big\},
\label{eq:current}
\end{multline}
or, equivalently, by the charge, magnetic and quadrupole form factors
\begin{gather}
G_C=G_1+\tfrac23\eta G_Q,\qquad G_M=G_2,\nonumber\\
G_Q=G_1-G_2+(1+\eta)G_3,
\label{eq:GCMQ}
\end{gather}
with $\eta=Q^2/4M_V^2$. At $Q^2=0$, they give the charge, $G_C(0)=1$, the
magnetic moment, $G_M(0)=2M_V\mu_V$, and the quadrupole moment,
$G_Q(0)=M_V^2Q_V$, and the charge radius is
$\langle r_C^2\rangle=-6\,dG_C/dQ^2|_{Q^2=0}$. We work in the Drell--Yan frame, $q^+=0$ and
$\mathbf{q}_\perp=(Q,0)$, where the plus component of the current does not
change the quark helicities and the matrix elements
are~\cite{Choi:2004ww,Gurjar:2024wpq}
\begin{multline}
I_{\Lambda'\Lambda}(Q^2)\equiv\frac{\langle V(P',\Lambda')|J^+|V(P,\Lambda)\rangle}{2P^+}\\
=\sum_{h\bar h}\int\frac{dx\,d^2\mathbf{k}_\perp}{16\pi^3}\,
\Psi^{\Lambda'*}_{h\bar h}\big(x,\mathbf{k}_\perp+(1-x)\mathbf{q}_\perp\big)\,
\Psi^{\Lambda}_{h\bar h}(x,\mathbf{k}_\perp),
\label{eq:Iplus}
\end{multline}
The neutral $\rho^0$, $\phi$, $J/\psi$ and
$\psi(2S)$ are eigenstates of charge conjugation, and their electromagnetic form factors
vanish. For the $\rho$, we therefore present the form factors of the charged $\rho^+$,
which has the same LFWF in the isospin limit and is also the state studied on the
lattice, while for the $\phi$, $J/\psi$ and $\psi(2S)$, we use the current of a single
quark with its charge removed. Parity and time-reversal invariance
leave four independent amplitudes, $I_{++}$, $I_{+0}$, $I_{+-}$ and $I_{00}$,
for three form factors, and in a truncated Fock space, they do not satisfy
the angular condition exactly, so a prescription is needed. We use the
Grach--Kondratyuk prescription~\cite{Grach:1983hd}, which does not use
$I_{00}$, 
\begin{align}
G_C&=\frac13\Big[(3-2\eta)I_{++}+2\sqrt{2\eta}\,I_{+0}+I_{+-}\Big],\nonumber\\
G_M&=2\Big[I_{++}-\frac{I_{+0}}{\sqrt{2\eta}}\Big],\nonumber\\
G_Q&=\frac{1}{\eta}\Big[-\eta I_{++}+\sqrt{2\eta}\,I_{+0}-I_{+-}\Big].
\label{eq:GK}
\end{align}

We use Eq.~(\ref{eq:GK}) for all the form factors. Other prescriptions, such
as those of Brodsky and Hiller \cite{Brodsky:1992px}, Chung \textit{et al.}
\cite{Chung:1988my} and Frankfurt \textit{et al.} \cite{Frankfurt:1993ut}, can also be
used. They give the same static moments and differ at finite $Q^2$ by an amount set by
the angular condition.

\par
\begin{figure*}[t]
\centering
\includegraphics[width=\textwidth]{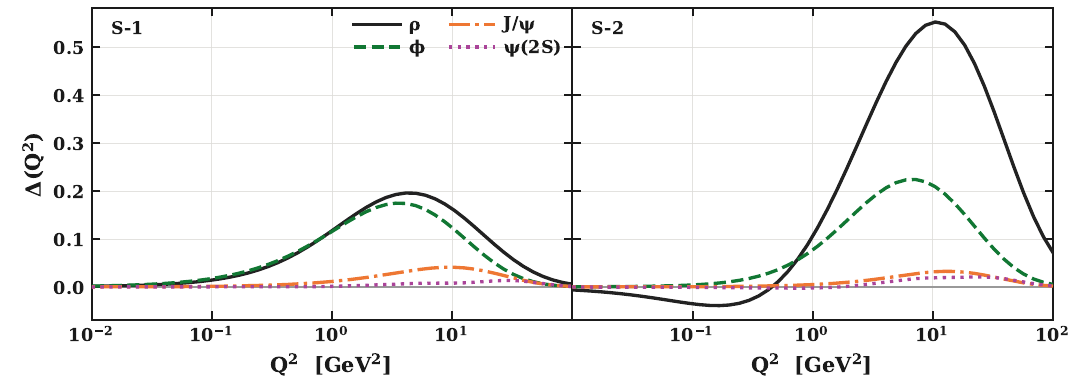}
\caption{(Color online) Angular condition $\Delta(Q^2)$ with S-1 (left) and S-2 (right)
for the $\rho$, $\phi$, $J/\psi$ and $\psi(2S)$ mesons.}
\label{fig:angcond}
\end{figure*}

 Since three form factors are described by four independent helicity
amplitudes, rotational covariance of the current requires one relation among
them, the angular condition~\cite{Grach:1983hd,Brodsky:1992px,Choi:2004ww}
\begin{equation}
\Delta(Q^2)=(1+2\eta)I_{++}-\sqrt{8\eta}\,I_{+0}+I_{+-}-I_{00}=0 .
\label{eq:angcond}
\end{equation}
We have checked that Eq.~(\ref{eq:angcond}) is satisfied identically by the
covariant decomposition, Eq.~(\ref{eq:current}), in our phase convention. In
the truncated $q\bar q$ Fock space the plus component of the current is not
rotationally covariant, and $\Delta(Q^2)$ therefore measures the violation of
rotational symmetry caused by the truncation and by the choice of spin
structure. Figure~\ref{fig:angcond} shows $\Delta(Q^2)$ for the four mesons with S-1 and S-2.
At $Q^2=0$, $\Delta$ vanishes, since all the amplitudes are fixed there by the charge
normalization. For S-1,
$\Delta(Q^2)$ grows smoothly with $Q^2$, reaches a maximum of about 0.20 for the
$\rho$ at $Q^2\approx4$~GeV$^2$ and 0.18 for the $\phi$ at $Q^2\approx3$~GeV$^2$,
and vanishes again at large $Q^2$ together with the amplitudes themselves. S-2 behaves
differently. For the $\rho$, $\Delta(Q^2)$ is small and
slightly negative below $Q^2\approx0.5$~GeV$^2$, but it then rises rapidly and
reaches about 0.55 at $Q^2\approx10$~GeV$^2$, i.e., about three times the
maximum violation found with S-1. For the $\phi$, the maximum violation of S-2,
about 0.22 at $Q^2\approx7$~GeV$^2$, is closer to that of S-1, because the two spin structures
converge as the quark mass increases. The Melosh--Wigner rotated S-1, which is unitary at every
point of $(x,\mathbf{k}_\perp)$, thus preserves rotational covariance considerably better than S-2
for $Q^2\gtrsim1$~GeV$^2$, which is where the form factors are compared with lattice QCD. The Grach--Kondratyuk prescription
does not use $I_{00}$, which is known to be the amplitude most sensitive to the
contributions beyond the valence Fock state~\cite{Choi:2004ww}. For charmonium the violation is small for both spin structures, with a maximum of about 0.04
for the $J/\psi$ and 0.02 for the $\psi(2S)$, as relativistic effects become weaker for heavy
quarks.

\par
\begin{figure*}[t]
\centering
\includegraphics[width=\textwidth]{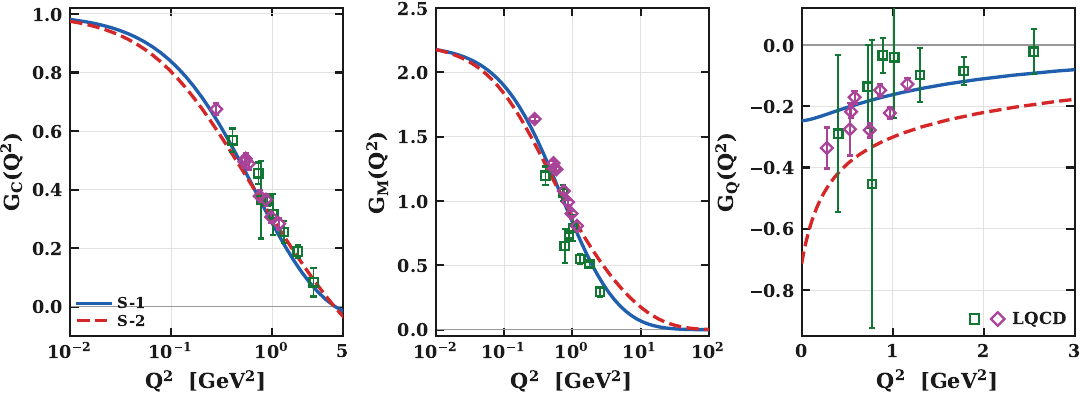}
\caption{(Color online) Charge (left), magnetic (middle) and quadrupole
(right) form factors of the $\rho^+$ with S-1 (solid blue) and S-2 (dashed
red), obtained with the Grach--Kondratyuk prescription, Eq.~(\ref{eq:GK}).
The LQCD results for the $\rho^+$ are from
QCDSF~\cite{QCDSF:2008tjq} at $m_\pi=406$~MeV (squares) and the Hadron
Spectrum collaboration~\cite{Shultz:2015pfa} at $m_\pi\approx700$~MeV
(diamonds). For the latter, points obtained with different operators at the
same $Q^2$ are averaged. $G_C$ is shown for $10^{-2}\le Q^2\le5$~GeV$^2$ and
$G_M$ for $10^{-2}\le Q^2\le10^{2}$~GeV$^2$ on a logarithmic scale.}
\label{fig:ff}
\end{figure*}

\begin{figure}[!t]
\centering
\includegraphics[width=\columnwidth]{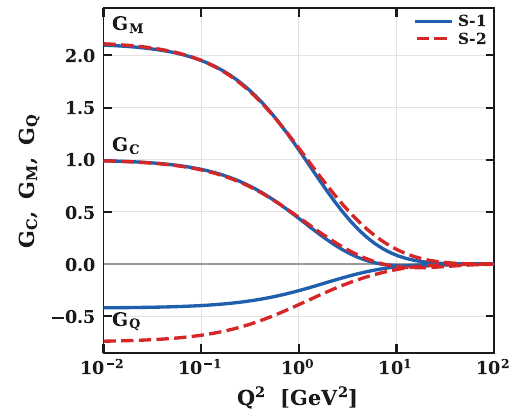}
\caption{(Color online) Charge, magnetic and quadrupole form factors of the
$\phi$ meson with S-1 (solid blue) and S-2 (dashed red), obtained with the
Grach--Kondratyuk prescription.}
\label{fig:ffphi}
\end{figure}

 The electromagnetic form factors test the same wave functions in a process that does not
involve the proton at all. Figure~\ref{fig:ff} shows the charge, magnetic and quadrupole form
factors of the $\rho^+$ obtained from Eqs.~(\ref{eq:Iplus}) and (\ref{eq:GK}), together with the
lattice QCD results of the QCDSF collaboration at $m_\pi=406$~MeV~\cite{QCDSF:2008tjq} and the
Hadron Spectrum collaboration at $m_\pi\approx700$~MeV~\cite{Shultz:2015pfa}. With S-1, the
$\rho^+$ results follow the trend of the lattice data. At $Q^2\simeq0.86$~GeV$^2$, for example,
the Hadron Spectrum values are $G_C=0.37$, $G_M=0.99$ and $G_Q=-0.15$, and S-1 gives 0.33, 0.92 and
$-0.17$. At the lowest lattice $Q^2$, the lattice values of $G_C$ and $G_M$ lie slightly above
our $\rho^+$ results, towards the $\phi$ results shown in Fig.~\ref{fig:ffphi}.
This is expected, since the lattice calculations use
heavier than physical quark masses, where the $\rho$ meson is heavier
($m_\rho\approx0.90$ and 1.02~GeV) and more compact than the physical one. At
larger $Q^2$, the charge form factor changes its sign at $Q^2\simeq4$~GeV$^2$
for the $\rho$ and at 6--7~GeV$^2$ for the $\phi$. The $G_C$ and $G_M$
obtained with S-1 and S-2 are similar, whereas $G_Q$ is significantly
different, with S-2 giving a much larger $|G_Q|$ due to its different
longitudinal component. Above $Q^2\approx0.8$~GeV$^2$, most of the lattice
values of $G_Q$ lie close to S-1, although with sizeable scatter, while S-2 is
about a factor of two larger in magnitude. The $\phi$ form factors, shown in
Fig.~\ref{fig:ffphi}, fall more slowly than those of the $\rho$, as expected for the more
compact $s\bar s$ state, and again only $G_Q$ is sensitive to the spin structure. Among the static
properties in Table~\ref{tab:static}, the $\rho^+$ charge radius is 0.67~fm with S-1 and 0.78~fm
with S-2, while that of the $\phi$ is about 0.5~fm for both. The magnetic moments,
$G_M(0)=2.12$--$2.23$, are close to the HadSpec lattice value 2.17(10)~\cite{Shultz:2015pfa} and show a weak
dependence on the spin structure. The quadrupole moment of the $\rho^+$, by contrast, changes
by a factor of about three between S-1 and S-2. The
lattice value $G_Q(0)=-0.54(10)$ \cite{Shultz:2015pfa}, obtained by extrapolating from finite $Q^2$,
lies between the two. A precise lattice determination of the $\rho^+$ quadrupole form factor at
small $Q^2$ and physical quark masses would therefore provide an independent test of the spin
structure.

\par
\begin{figure*}[t]
\centering
\includegraphics[width=\textwidth]{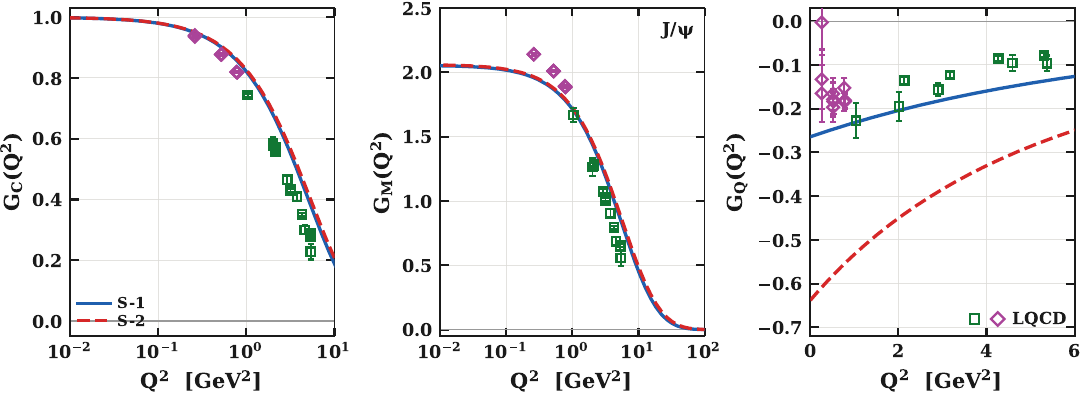}
\caption{(Color online) Charge (left), magnetic (middle) and quadrupole (right) form
factors of the $J/\psi$ with S-1 (solid blue) and S-2 (dashed red), compared with the
quenched (squares)~\cite{Dudek:2006ej} and $N_f=2+1$ (diamonds)~\cite{Delaney:2023fsx}
LQCD results.}
\label{fig:ffjpsi}
\end{figure*}

 Figure~\ref{fig:ffjpsi} compares the $J/\psi$ form factors with the lattice QCD
results~\cite{Dudek:2006ej,Delaney:2023fsx}. The $G_C$
and $G_M$ of S-1 and S-2 are almost the same, as the spin rotation is small for heavy
quarks. At small $Q^2$, our results are close to the $N_f=2+1$ lattice data \cite{Delaney:2023fsx}.
At $Q^2\simeq0.8$~GeV$^2$, for example, S-1 gives $G_C=0.85$ and $G_M=1.78$, compared to 0.82 and
1.89.
At larger $Q^2$, our form factors fall more slowly than the quenched lattice data,
which corresponds to a smaller charge radius, 0.22~fm compared to 0.25--0.26~fm \cite{Dudek:2006ej,Delaney:2023fsx}. The
magnetic moment, $G_M(0)=2.06$, is close to the quenched value 2.10(3) \cite{Dudek:2006ej}. Again, $G_Q$ is
the most sensitive to the spin structure. S-1 gives $G_Q(0)=-0.26$, close to the lattice
values $-0.23(2)$ \cite{Dudek:2006ej} and $-0.172(15)$ \cite{Delaney:2023fsx}, while S-2 gives $-0.64$. Since the $S$-wave LFWF has no
$D$-wave component, the quadrupole moment here comes entirely from the spin structure.

%======================================================================
\section{Summary and conclusions}
\label{sec:summary}
%======================================================================

\par We have studied exclusive diffractive production of the $\rho^0$, $\phi$, $J/\psi$ and
$\psi(2S)$ in the color-dipole framework with two spin-orbit light-front wave functions, the
Melosh--Wigner rotated S-1 of the light-front quark model and the S-2 form used in holographic and
boosted-Gaussian dipole studies. The calculations have been compared with all the \nPoints\ data
points of H1, ZEUS, E665, NMC and HERMES for the $\rho^0$ and $\phi$, and with the H1 and ZEUS data
on $\sigma_{\psi(2S)}/\sigma_{J/\psi}$.

\par Both spin structures describe the $Q^2$, $W$ and $t$ dependences of the cross sections well
and differ mainly by an overall factor. With the dipole cross section fitted to the
structure-function data with the LFQM quark masses, S-1 reproduces the normalization of the HERA
data, whereas S-2 overshoots it by a factor of 1.5--1.8 for the H1 and ZEUS $\rho^0$ and the H1
$\phi$ data. The normalization, however, also depends on the skewedness correction, which is not
included, and on the diffractive slope. The observables in which these factors cancel favor S-1.
For $R=\sigma_L/\sigma_T$, S-1 gives $\chi^2/N=\chiTwoSOne$ against \chiTwoSTwo\ for S-2 over the
HERA data, and the decay constants, which do not involve the dipole model at all, agree with the
data within about 6\% for S-1, whereas S-2 overestimates them by more than 20\% for the $\rho$ and
$\phi$. The ratio $\sigma_\phi/\sigma_\rho$ is described by both. The difference between the two
spin structures in $R$ drops from about 90\% for the $\rho$ to about 45\% for the $\phi$, as
expected for a relativistic effect that weakens with increasing quark mass. The $\rho^+$ form
factors are consistent with the lattice QCD results once the heavier lattice masses are taken into
account, and the quadrupole moment is the static quantity most sensitive to the spin structure.
The light-front angular condition is violated much less with S-1 than with S-2 for the $\rho$ at
$Q^2\gtrsim1$~GeV$^2$. Precise measurements of $R$ for light vector mesons at the
EIC~\cite{AccardiEIC,EICYellowReport} can therefore test the spin structure of the vector-meson
LFWF directly.

\par For the $J/\psi$ and $\psi(2S)$, S-2 describes the measured ratio
$\sigma_{\psi(2S)}/\sigma_{J/\psi}$ as a function of $Q^2$ and $W$, while S-1 lies below the data.
S-2 also agrees with the measured $J/\psi$ decay constant and leptonic width, whereas S-1 is about
9\% lower. The lattice quadrupole form factor of the $J/\psi$, on the other hand, lies closer to
S-1. While the light-meson observables favor S-1, the charmonium ratio and the $J/\psi$ decay
constant thus favor S-2. Since the ratio depends on the node of the $\psi(2S)$, a better
description of the radially excited state, for example through $1S$--$2S$ mixing, will be
important. The violation of the angular condition is small for charmonium.

 It will be interesting to extend this study to other dipole
models~\cite{Kowalski:2003hm,Watt:2007nr,Rezaeian2013}, to include the skewedness correction, and
to use radial wave functions from holographic QCD.

\begin{table}[!t]
\caption{Normalization constants of the S-2 wave function, Eq.~(\ref{eq:NLNT}). For
S-1, $\mathcal{N}_L=\mathcal{N}_T=1$.}
\label{tab:norm}
\begin{ruledtabular}
\begin{tabular}{lcc}
 & $\mathcal{N}_L$ & $\mathcal{N}_T$ [GeV$^2$] \\
\hline
$\rho$ & 3.509 & 1.097 \\
$\phi$ & 3.986 & 1.938 \\
$J/\psi$ & 2.783 & 13.07 \\
$\psi(2S)$ & 2.092 & 13.98 \\
\end{tabular}
\end{ruledtabular}
\end{table}
\section{Appendix}
\appendix
\section{\texorpdfstring{Normalization of the S-2 wave function}{Normalization of the S-2 wave function}}
\label{app:norm}

\par Since the S-2 spin-orbit wave function of Eq.~(\ref{eq:S2mat}) is not unitary, the
longitudinal and transverse states are normalized separately with
\begin{align}
\mathcal{N}_L&=\frac12\int\frac{dx\,d^2\mathbf{k}_\perp}{16\pi^3}\,\phi^2
\left[1+\frac{M_0^2}{M_V^2}\right]^2,\nonumber\\
\mathcal{N}_T&=\int\frac{dx\,d^2\mathbf{k}_\perp}{16\pi^3}\,\phi^2\,
\frac{m_q^2+\left[x^2+(1-x)^2\right]\mathbf{k}_\perp^2}{4x^2(1-x)^2},
\label{eq:NLNT}
\end{align}
which are related to $n_\Lambda$ of Eq.~(\ref{eq:hankel}) by $n_{0}=4\pi\mathcal{N}_L$ and
$n_{\pm1}=4\pi\mathcal{N}_T$. For S-1, both are equal to one because of
Eq.~(\ref{eq:unit}). The values for the four mesons are listed in Table~\ref{tab:norm}.

\section{\texorpdfstring{Variational determination of $m_q$ and $\beta$}{Variational determination of m and beta}}
\label{app:var}

\begin{table}[!t]
\caption{Potential parameters of Eq.~(\ref{eq:Vparts}).}
\label{tab:pot}
\begin{ruledtabular}
\begin{tabular}{lcccc}
 & $a$ [GeV] & $b$ & $c$ [GeV] & $\alpha_s$ \\
\hline
$\rho$, $\phi$~\cite{ChoiJi1999,Ji:1998zi} & $-0.144$ & 0.010329~GeV$^3$ & --- & 0.607 \\
$J/\psi$, $\psi(2S)$~\cite{Ridwan:2025} & $-0.411$ & 0.18~GeV$^2$ & 0.027 & 0.402 \\
\end{tabular}
\end{ruledtabular}
\end{table}

\par For all four mesons, the mass eigenvalue equation is
$H_{q\bar q}|\Psi\rangle=(H_0+V_{q\bar q})|\Psi\rangle=M|\Psi\rangle$, with the kinetic term
$H_0=2\sqrt{m_q^2+\mathbf{k}^2}$ and the interaction
\begin{equation}
V_{q\bar q}=V_{\rm conf}+V_{\rm Coul}+V_{\rm hyp},
\label{eq:HHO}
\end{equation}
where
\begin{align}
V_{\rm conf}&=a+br^2\quad(\rho,\ \phi),\nonumber\\
V_{\rm conf}&=a+\frac bc\left(1-e^{-cr}\right)\quad(J/\psi,\ \psi(2S)),\nonumber\\
V_{\rm Coul}&=-\frac{4\alpha_s}{3r},\nonumber\\
V_{\rm hyp}&=\frac{32\pi\alpha_s}{9m_q^2}\,\langle\mathbf{S}_q\cdot\mathbf{S}_{\bar q}\rangle\,\delta^3(\mathbf{r}),
\label{eq:Vparts}
\end{align}
with $\langle\mathbf{S}_q\cdot\mathbf{S}_{\bar q}\rangle=1/4$ for a vector meson. The potential
parameters are listed in Table~\ref{tab:pot}. With the Gaussian wave function of
Eq.~(\ref{eq:radial}), all the expectation values are analytic.
With $z=m_q^2/\beta^2$ and $u=c/2\beta$, they are
\begingroup\allowdisplaybreaks
\begin{align}
\langle H_0\rangle&=\frac{2\beta}{\sqrt\pi}\,z\,e^{z/2}K_1\!\left(\frac z2\right),\nonumber\\
\langle V_{\rm conf}\rangle&=a+\frac{3b}{2\beta^2}\quad(\rho,\ \phi),\nonumber\\
\langle V_{\rm conf}\rangle&=a+\frac bc\left[1-(1+2u^2)\,e^{u^2}\mathrm{erfc}(u)+\frac{2u}{\sqrt\pi}\right]\nonumber\\
&\hspace{4.2cm}(J/\psi,\ \psi(2S)),\nonumber\\
\langle V_{\rm Coul}\rangle&=-\frac{8\alpha_s\beta}{3\sqrt\pi},\nonumber\\
\langle V_{\rm hyp}\rangle&=\frac{32\alpha_s\beta^3}{9\sqrt\pi\,m_q^2}\,
\langle\mathbf{S}_q\cdot\mathbf{S}_{\bar q}\rangle,
\label{eq:Hexp}
\end{align}
\endgroup
For a given $m_q$, the Gaussian parameter follows from the variational condition
\begin{equation}
\frac{\partial\langle H_0+V_{\rm conf}+V_{\rm Coul}\rangle}{\partial\beta}=0 ,
\label{eq:varcond}
\end{equation}
in which the hyperfine term is treated as a perturbation. The vector-meson mass is then
$M_V=\langle H_0+V_{\rm conf}+V_{\rm Coul}+V_{\rm hyp}\rangle$, and $m_q$ is obtained with the
physical masses of the $\rho^0$, $\phi$ and $J/\psi$, 775.26, 1019.46 and
3096.9~MeV~\cite{PDG2024}. The resulting $m_q$ and $\beta$ are listed in Table~\ref{tab:par}.
For the $\rho$, $M_V$ has a minimum of 768~MeV at $m_{u,d}\approx0.235$~GeV as a function of
$m_{u,d}$, so that two masses give 775.26~MeV. The smaller one, $m_{u,d}=0.208$~GeV, gives a
negative pion mass and is discarded. For charmonium, the same Hamiltonian gives
$M_{\eta_c}=3.007$~GeV, 23~MeV above experiment, and, with the $2S$ function of
Eq.~(\ref{eq:radial2S}) and the same $m_c$ and $\beta$, $M_{\psi(2S)}=3.6483$~GeV, 38~MeV below
experiment. The harmonic-oscillator potential of the $\rho$ and $\phi$ could
also be used for charmonium. It gives $m_c=1.712$~GeV and $\beta=0.675$~GeV, but its $2S$
wave function overestimates the $\psi(2S)$ decay constant by 37\% and 64\% (403 and 482~MeV with
S-1 and S-2, compared to 294~MeV \cite{PDG2024}), i.e.\ the $\psi(2S)$ wave function at
small distances, to which the ratio $\sigma_{\psi(2S)}/\sigma_{J/\psi}$ is sensitive. For
the $J/\psi$ and $\psi(2S)$, we therefore use the screened potential of
Ref.~\cite{Ridwan:2025}, which gives 246 and 301~MeV (Table~\ref{tab:static}).

\section{Fit of the dipole cross section}
\label{app:cgc}

\par The parameters of the dipole cross section, Eq.~(\ref{eq:cgc}), are fitted to deep-inelastic
structure-function data with the quark masses of Table~\ref{tab:par}, which are also used in the
photon wave function of the vector-meson amplitude. Through the optical theorem, the cross
sections of a transversely and a longitudinally polarized virtual photon on the proton
are~\cite{Golec-Biernat:1998zce,KowalskiMotykaWatt2006,AhmadyEtAl2016}
\begin{multline}
\sigma^{\gamma^*p}_{T,L}(x_{Bj},Q^2)=\sum_f\int d^2\mathbf{r}_\perp\int_0^1\frac{dz}{4\pi}\\
\times\big(\Psi^*\Psi\big)^f_{T,L}(z,r_\perp;Q^2)\,\hat\sigma(x^f_m,r_\perp),
\label{eq:sigLTdis}
\end{multline}
where $z$ is the light-front momentum fraction of the quark, and the photon wave functions of
Eq.~(\ref{eq:photon}), summed over the helicities, give
\begin{align}
\big(\Psi^*\Psi\big)^f_T&=\frac{2N_c}{\pi}\,\alpha_{\rm em}e_f^2\Big\{\big[z^2+(1-z)^2\big]
\varepsilon_f^2K_1^2(\varepsilon_fr_\perp)\nonumber\\
&\qquad+m_f^2K_0^2(\varepsilon_fr_\perp)\Big\},\nonumber\\
\big(\Psi^*\Psi\big)^f_L&=\frac{8N_c}{\pi}\,\alpha_{\rm em}e_f^2\,Q^2z^2(1-z)^2
K_0^2(\varepsilon_fr_\perp),
\label{eq:photonsq}
\end{align}
with $\varepsilon_f^2=z(1-z)Q^2+m_f^2$. As in Refs.~\cite{Golec-Biernat:1998zce,AhmadyEtAl2016}, the
dipole cross section is evaluated at $x^f_m=x_{Bj}(1+4m_f^2/Q^2)$ for every flavor. The sum runs
over $f=u,d,s,c$ with $m_u=m_d=0.266$, $m_s=0.497$ and $m_c=1.606$~GeV. The structure functions are
$F_2=Q^2(\sigma^{\gamma^*p}_T+\sigma^{\gamma^*p}_L)/(4\pi^2\alpha_{\rm em})$ and
$F_L=Q^2\sigma^{\gamma^*p}_L/(4\pi^2\alpha_{\rm em})$~\cite{Golec-Biernat:1998zce,AhmadyEtAl2016},
and the reduced cross section measured at HERA is, for $Q^2\ll M_Z^2$~\cite{H1ZEUS:2015HERA},
\begin{equation}
\sigma_r=F_2-\frac{y^2}{1+(1-y)^2}\,F_L,\qquad y=\frac{Q^2}{s\,x_{Bj}},
\label{eq:sigr}
\end{equation}
with $\sqrt s$ the $ep$ center-of-mass energy. For the charm reduced cross section only the charm
term of Eq.~(\ref{eq:sigLTdis}) is kept, and the E665 and NMC data, which are quoted as $F_2$, are
compared with $F_2$ directly.

\par We fit all the structure-function data with $x_{Bj}\le0.01$ and $0.045\le Q^2\le45$~GeV$^2$,
namely the combined H1 and ZEUS inclusive reduced cross sections~\cite{H1ZEUS:2015HERA} at $\sqrt s=318$,
300, 251 and 225~GeV (524 points), the combined charm reduced cross sections~\cite{H1:2018flt} at
$\sqrt s=318$~GeV (34 points), and the proton $F_2$ of E665~\cite{E665:1996mob} (45 points) and
NMC~\cite{NewMuon:1996fwh} (9 points), 612 points in all. The H1 measurement of
$F_L$~\cite{H1:2013ktq} is not included, since it is partly based on the same events as the
combined data. We minimize
\begin{equation}
\chi^2=\sum_{i=1}^{N}\frac{\big[F^{\rm th}_i-F_i\big]^2}{\delta_i^2},
\label{eq:chi2cgc}
\end{equation}
where $F_i$ is the measured $\sigma_r$ or $F_2$ and $\delta_i$ its total uncertainty (statistical,
systematic and procedural uncertainties added in quadrature). As in
Refs.~\cite{IancuItakuraMunier2004,AhmadyEtAl2016}, $\mathcal{N}_0=0.7$ and $\kappa_0=9.9$ are
fixed, and $\sigma_0$, $\gamma_s$, $\lambda$ and $x_0$ are free. The minimization uses the
least-squares algorithm in SciPy~\cite{Virtanen:2019joe}. As a check, the
masses and parameters of fit A of Ref.~\cite{AhmadyEtAl2016} give $\chi^2=535.6$ for the 524 HERA
inclusive points, compared with 535 quoted there.

\begin{table}[!t]
\caption{Fitted parameters of the dipole cross section and $\chi^2/N$ of the fit for each data set
and in total ($\chi^2/\mathrm{dof}=1719.3/608=2.83$). The uncertainties are those of
$\Delta\chi^2=1$ multiplied by $\sqrt{\chi^2/\mathrm{dof}}$.}
\label{tab:cgc}
\begin{ruledtabular}
\begin{tabular}{lcc}
$m_{u,d}$, $m_s$, $m_c$ [GeV] & \multicolumn{2}{c}{0.266, 0.497, 1.606} \\
$\sigma_0$ [mb] & \multicolumn{2}{c}{$58.9\pm3.5$} \\
$\gamma_s$ & \multicolumn{2}{c}{$0.740\pm0.020$} \\
$\lambda$ & \multicolumn{2}{c}{$0.161\pm0.006$} \\
$x_0$ & \multicolumn{2}{c}{$\big(1.16^{+1.82}_{-0.71}\big)\times10^{-7}$} \\
$\mathcal{N}_0$, $\kappa_0$ & \multicolumn{2}{c}{0.7, 9.9 (fixed)} \\
\hline
data & $N$ & $\chi^2/N$ \\
\hline
HERA $\sigma_r$~\cite{H1ZEUS:2015HERA} & 524 & 2.47 \\
HERA $\sigma_r^{c\bar c}$~\cite{H1:2018flt} & 34 & 7.64 \\
E665 $F_2^p$~\cite{E665:1996mob} & 45 & 2.67 \\
NMC $F_2^p$~\cite{NewMuon:1996fwh} & 9 & 5.17 \\
\hline
total & 612 & 2.81 \\
\end{tabular}
\end{ruledtabular}
\end{table}

\par
\begin{figure*}[p]
\centering
\includegraphics[width=\textwidth]{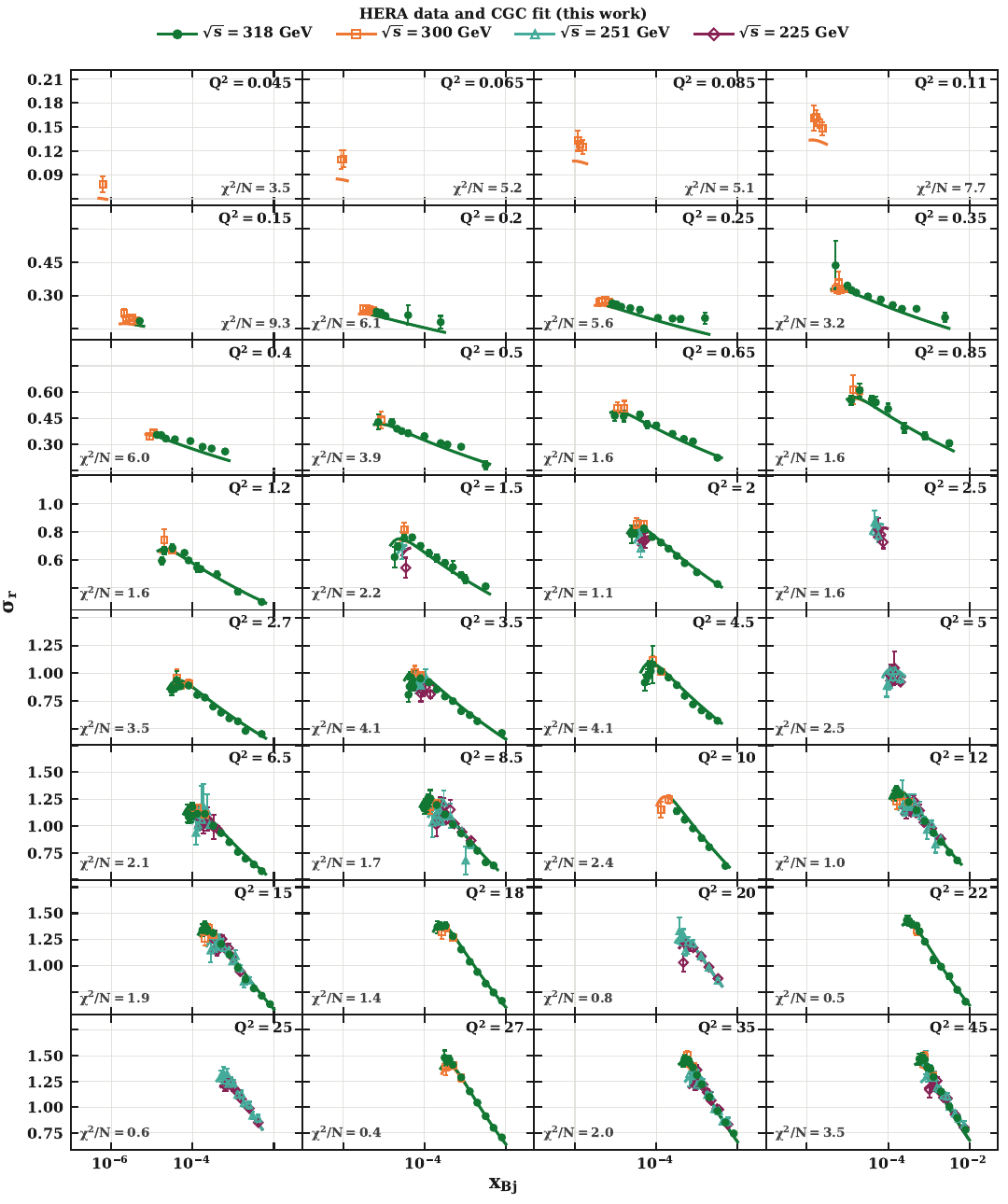}
\caption{HERA reduced cross section $\sigma_r$~\cite{H1ZEUS:2015HERA} as a function of $x_{Bj}$ in
the 32 $Q^2$ bins (in GeV$^2$) of the fit, compared with the fitted dipole cross section (lines).
The fit is calculated at the $\sqrt s$ of the data and drawn in the same color as the data of that
energy. The $\chi^2/N$ of each bin is given in the panels, and the $y$ axis is common to the panels
of each row.}
\label{fig:cgcfit}
\end{figure*}

\begin{figure*}[t]
\centering
\includegraphics[width=\textwidth]{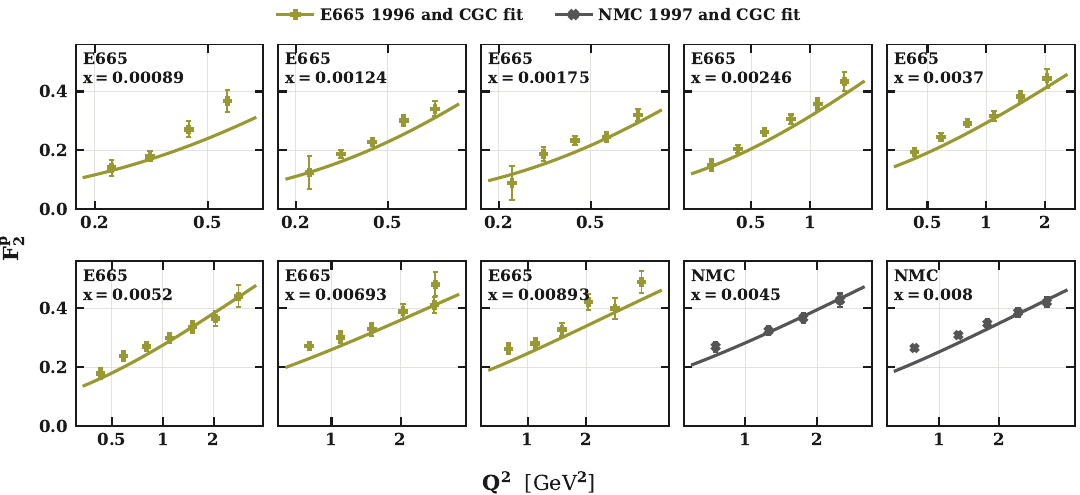}
\caption{Proton structure function $F_2^p$ at $x_{Bj}\le0.01$ measured by E665~\cite{E665:1996mob}
and NMC~\cite{NewMuon:1996fwh}, compared with the fit (lines, in the color of the data).}
\label{fig:cgcF2}
\end{figure*}

\begin{figure*}[t]
\centering
\includegraphics[width=\textwidth]{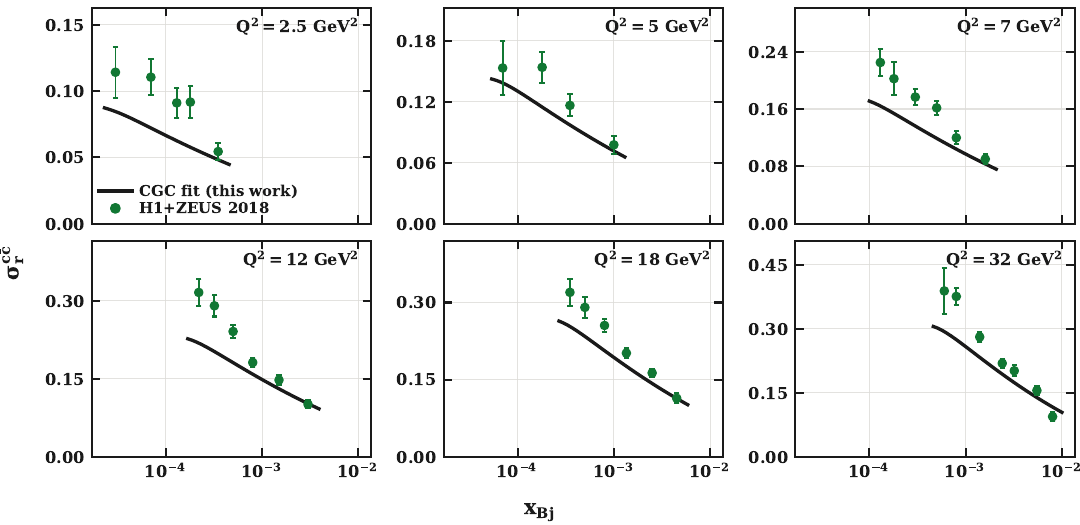}
\caption{Charm reduced cross section measured by H1 and ZEUS~\cite{H1:2018flt} at $\sqrt s=318$~GeV,
compared with the fit (solid lines).}
\label{fig:cgccharm}
\end{figure*}

 The fitted parameters and the $\chi^2/N$ of each data set are listed in Table~\ref{tab:cgc}.
The parameters are strongly correlated (correlation coefficients 0.92--0.99 in magnitude) and
must be used as a set. The fit is compared with the data in Figs.~\ref{fig:cgcfit}--\ref{fig:cgccharm}.
The HERA inclusive data are described with $\chi^2/N=2.47$, compared with about 1.0 for fits with
light quark masses~\cite{AhmadyEtAl2016}. The deviations come mainly from $Q^2<1$~GeV$^2$, where the
calculation lies below the data ($\chi^2/N=4.5$ for these 94 points, against 2.0 for the 430 points
with $Q^2\ge1$~GeV$^2$). A similar deterioration of the fit with a constituent-size light-quark mass
was found in the GBW model, where $m_l=0.28$~GeV gives $\chi^2/N_{\rm dof}=2.78$~\cite{Golec-Biernat:2017lfv}. The E665 and NMC data, at $W\approx9$--26~GeV, lie on average 8\% and 5\%
above the fit, again mostly at the lowest $Q^2$. The charm data are underestimated by 12--28\% in
the six $Q^2$ bins, which reflects the large charm mass of the $J/\psi$ wave function. Evaluating
the dipole cross section at $x_{Bj}$ for the light quarks and shifting only the charm
term~\cite{KowalskiMotykaWatt2006} would lower the total $\chi^2/N$ to 2.0 and change $\sigma_0$
by about 10\%.

\section*{Acknowledgments}
 The author acknowledges the use of ChatGPT for grammar checking and language editing of the manuscript.

\FloatBarrier
\bibliography{refs_prd}

@article{Sharma:2023njj,
    author = "Sharma, Neetika",
    title = "{Exclusive diffractive J/{\ensuremath{\psi}} and {\ensuremath{\psi}}(2S) production in dipole model using a holographic AdS/QCD light-front wavefunction with longitudinal confinement}",
    eprint = "2311.09152",
    archivePrefix = "arXiv",
    primaryClass = "hep-ph",
    doi = "10.1103/PhysRevD.109.014019",
    journal = "Phys. Rev. D",
    volume = "109",
    number = "1",
    pages = "014019",
    year = "2024"
}

@article{Choi:1997iq,
    author = "Choi, Ho-Meoyng and Ji, Chueng-Ryong",
    title = "{Mixing angles and electromagnetic properties of ground state pseudoscalar and vector meson nonets in the light cone quark model}",
    eprint = "hep-ph/9711450",
    archivePrefix = "arXiv",
    reportNumber = "NCSU-97-32",
    doi = "10.1103/PhysRevD.59.074015",
    journal = "Phys. Rev. D",
    volume = "59",
    pages = "074015",
    year = "1999"
}

@article{Ridwan:2024ngc,
    author = "Ridwan, Muhammad and Arifi, Ahmad Jafar and Mart, Terry",
    title = "{Self-consistent M1 radiative transitions of excited Bc and heavy quarkonia with different polarizations in the light-front quark model}",
    eprint = "2409.13172",
    archivePrefix = "arXiv",
    primaryClass = "hep-ph",
    doi = "10.1103/PhysRevD.111.016011",
    journal = "Phys. Rev. D",
    volume = "111",
    number = "1",
    pages = "016011",
    year = "2025"
}

@article{AccardiEIC,
    author = "Accardi, A. and others",
    editor = "Deshpande, A. and Meziani, Z. E. and Qiu, J. W.",
    title = "{Electron Ion Collider: The Next QCD Frontier}: {Understanding the glue that binds us all}",
    eprint = "1212.1701",
    archivePrefix = "arXiv",
    primaryClass = "nucl-ex",
    reportNumber = "BNL-98815-2012-JA, JLAB-PHY-12-1652",
    doi = "10.1140/epja/i2016-16268-9",
    journal = "Eur. Phys. J. A",
    volume = "52",
    number = "9",
    pages = "268",
    year = "2016"
}

@article{Forshaw:2003ki,
    author = "Forshaw, Jeffrey R. and Sandapen, R. and Shaw, Graham",
    title = "{Color dipoles and rho, phi electroproduction}",
    eprint = "hep-ph/0312172",
    archivePrefix = "arXiv",
    doi = "10.1103/PhysRevD.69.094013",
    journal = "Phys. Rev. D",
    volume = "69",
    pages = "094013",
    year = "2004"
}

@article{EICYellowReport,
    author = "Abdul Khalek, R. and others",
    title = "{Science Requirements and Detector Concepts for the Electron-Ion Collider}: {EIC Yellow Report}",
    eprint = "2103.05419",
    archivePrefix = "arXiv",
    primaryClass = "physics.ins-det",
    reportNumber = "BNL-220990-2021-FORE, JLAB-PHY-21-3198, LA-UR-21-20953",
    doi = "10.1016/j.nuclphysa.2022.122447",
    journal = "Nucl. Phys. A",
    volume = "1026",
    pages = "122447",
    year = "2022"
}

@article{Acharyya:2024enp,
    author = "Acharyya, Ritwik and Puhan, Satyajit and Dahiya, Harleen",
    title = "{Quark spin-orbit correlations in spin-0 and spin-1 mesons using the light-front quark model}",
    eprint = "2405.00446",
    archivePrefix = "arXiv",
    primaryClass = "hep-ph",
    doi = "10.1103/PhysRevD.110.034020",
    journal = "Phys. Rev. D",
    volume = "110",
    number = "3",
    pages = "034020",
    year = "2024"
}

@article{AhmadyEtAl2016,
    author = "Ahmady, Mohammad and Sandapen, Ruben and Sharma, Neetika",
    title = "{Diffractive $\rho$ and $\phi$ production at HERA using a holographic AdS/QCD light-front meson wave function}",
    eprint = "1605.07665",
    archivePrefix = "arXiv",
    primaryClass = "hep-ph",
    doi = "10.1103/PhysRevD.94.074018",
    journal = "Phys. Rev. D",
    volume = "94",
    number = "7",
    pages = "074018",
    year = "2016"
}

@article{Arifi:2025olq,
    author = "Arifi, Ahmad Jafar and Choi, Ho-Meoyng and Ji, Chueng-Ryong",
    title = "{Beyond leading twist: {\ensuremath{\rho}} meson decay constants and distribution amplitudes in a self-consistent light-front quark model}",
    eprint = "2506.02844",
    archivePrefix = "arXiv",
    primaryClass = "hep-ph",
    doi = "10.1103/hhk8-g2bj",
    journal = "Phys. Rev. D",
    volume = "112",
    number = "3",
    pages = "033009",
    year = "2025"
}

@article{Arifi:2022qnd,
    author = "Arifi, Ahmad Jafar and Choi, Ho-Meoyng and Ji, Chueng-Ryong and Oh, Yongseok",
    title = "{Independence of current components, polarization vectors, and reference frames in the light-front quark model analysis of meson decay constants}",
    eprint = "2210.12780",
    archivePrefix = "arXiv",
    primaryClass = "hep-ph",
    doi = "10.1103/PhysRevD.107.053003",
    journal = "Phys. Rev. D",
    volume = "107",
    number = "5",
    pages = "053003",
    year = "2023"
}

@article{Armesto:2014sma,
    author = "Armesto, N{\'e}stor and Rezaeian, Amir H.",
    title = "{Exclusive vector meson production at high energies and gluon saturation}",
    eprint = "1402.4831",
    archivePrefix = "arXiv",
    primaryClass = "hep-ph",
    doi = "10.1103/PhysRevD.90.054003",
    journal = "Phys. Rev. D",
    volume = "90",
    number = "5",
    pages = "054003",
    year = "2014"
}

@article{Arnold:1979cg,
    author = "Arnold, R. G. and Carlson, Carl E. and Gross, Franz",
    title = "{Elastic electron-Deuteron Scattering at High-Energy}",
    reportNumber = "SLAC-PUB-2318",
    doi = "10.1103/PhysRevC.21.1426",
    journal = "Phys. Rev. C",
    volume = "21",
    pages = "1426",
    year = "1980"
}

@article{BakamjianThomas1953,
    author = "Bakamjian, B. and Thomas, L. H.",
    title = "{Relativistic particle dynamics. 2}",
    doi = "10.1103/PhysRev.92.1300",
    journal = "Phys. Rev.",
    volume = "92",
    pages = "1300--1310",
    year = "1953"
}

@article{Brodsky:1992px,
    author = "Brodsky, Stanley J. and Hiller, John R.",
    title = "{Universal properties of the electromagnetic interactions of spin one systems}",
    reportNumber = "SLAC-PUB-5763",
    doi = "10.1103/PhysRevD.46.2141",
    journal = "Phys. Rev. D",
    volume = "46",
    pages = "2141--2149",
    year = "1992"
}

@article{Brodsky:1994kf,
    author = "Brodsky, Stanley J. and Frankfurt, L. and Gunion, J. F. and Mueller, Alfred H. and Strikman, M.",
    title = "{Diffractive leptoproduction of vector mesons in QCD}",
    eprint = "hep-ph/9402283",
    archivePrefix = "arXiv",
    reportNumber = "SLAC-PUB-6412, CU-TP-617, UCD-93-36",
    doi = "10.1103/PhysRevD.50.3134",
    journal = "Phys. Rev. D",
    volume = "50",
    pages = "3134--3144",
    year = "1994"
}

@article{Brodsky:1997de,
    author = "Brodsky, Stanley J. and Pauli, Hans-Christian and Pinsky, Stephen S.",
    title = "{Quantum chromodynamics and other field theories on the light cone}",
    eprint = "hep-ph/9705477",
    archivePrefix = "arXiv",
    reportNumber = "SLAC-PUB-7484, MPIH-V1-1997",
    doi = "10.1016/S0370-1573(97)00089-6",
    journal = "Phys. Rept.",
    volume = "301",
    pages = "299--486",
    year = "1998"
}

@article{Brodsky:2014yha,
    author = "Brodsky, Stanley J. and de Teramond, Guy F. and Dosch, Hans Gunter and Erlich, Joshua",
    title = "{Light-Front Holographic QCD and Emerging Confinement}",
    eprint = "1407.8131",
    archivePrefix = "arXiv",
    primaryClass = "hep-ph",
    reportNumber = "SLAC-PUB-15972",
    doi = "10.1016/j.physrep.2015.05.001",
    journal = "Phys. Rept.",
    volume = "584",
    pages = "1--105",
    year = "2015"
}

@article{Choi:2004ww,
    author = "Choi, Ho-Meoyng and Ji, Chueng-Ryong",
    title = "{Electromagnetic structure of the rho meson in the light front quark model}",
    eprint = "hep-ph/0402114",
    archivePrefix = "arXiv",
    doi = "10.1103/PhysRevD.70.053015",
    journal = "Phys. Rev. D",
    volume = "70",
    pages = "053015",
    year = "2004"
}

@article{ChoiJi1999,
    author = "Choi, Ho-Meoyng and Ji, Chueng-Ryong",
    title = "{Mixing angles and electromagnetic properties of ground state pseudoscalar and vector meson nonets in the light cone quark model}",
    eprint = "hep-ph/9711450",
    archivePrefix = "arXiv",
    reportNumber = "NCSU-97-32",
    doi = "10.1103/PhysRevD.59.074015",
    journal = "Phys. Rev. D",
    volume = "59",
    pages = "074015",
    year = "1999"
}

@article{ChoiJi2007,
    author = "Choi, Ho-Meoyng and Ji, Chueng-Ryong",
    title = "{Distribution amplitudes and decay constants for (pi, K, rho, K*) mesons in light-front quark model}",
    eprint = "hep-ph/0701177",
    archivePrefix = "arXiv",
    doi = "10.1103/PhysRevD.75.034019",
    journal = "Phys. Rev. D",
    volume = "75",
    pages = "034019",
    year = "2007"
}

@article{E665rho1997,
    author = "Adams, M. R. and others",
    collaboration = "E665",
    title = "{Diffractive production of $\rho^0(770)$ mesons in muon proton interactions at 470-GeV}",
    reportNumber = "MPI-PHE-97-03, FERMILAB-PUB-97-103-E",
    doi = "10.1007/s002880050386",
    journal = "Z. Phys. C",
    volume = "74",
    pages = "237--261",
    year = "1997"
}

@article{ForshawSandapen2012,
    author = "Forshaw, J. R. and Sandapen, R.",
    title = "{An AdS/QCD holographic wavefunction for the rho meson and diffractive rho meson electroproduction}",
    eprint = "1203.6088",
    archivePrefix = "arXiv",
    primaryClass = "hep-ph",
    reportNumber = "MAN-HEP-2012-12",
    doi = "10.1103/PhysRevLett.109.081601",
    journal = "Phys. Rev. Lett.",
    volume = "109",
    pages = "081601",
    year = "2012"
}

@article{Frankfurt:1995jw,
    author = "Frankfurt, Leonid and Koepf, Werner and Strikman, Mark",
    title = "{Hard diffractive electroproduction of vector mesons in QCD}",
    eprint = "hep-ph/9509311",
    archivePrefix = "arXiv",
    reportNumber = "TAUP-2290-95",
    doi = "10.1103/PhysRevD.54.3194",
    journal = "Phys. Rev. D",
    volume = "54",
    pages = "3194--3215",
    year = "1996"
}

@article{Golec-Biernat:1998zce,
    author = "Golec-Biernat, Krzysztof J. and Wusthoff, M.",
    title = "{Saturation effects in deep inelastic scattering at low Q**2 and its implications on diffraction}",
    eprint = "hep-ph/9807513",
    archivePrefix = "arXiv",
    reportNumber = "DTP-98-50",
    doi = "10.1103/PhysRevD.59.014017",
    journal = "Phys. Rev. D",
    volume = "59",
    pages = "014017",
    year = "1998"
}

@article{Goloskokov:2005sd,
    author = "Goloskokov, S. V. and Kroll, P.",
    title = "{Vector meson electroproduction at small Bjorken-x and generalized parton distributions}",
    eprint = "hep-ph/0501242",
    archivePrefix = "arXiv",
    reportNumber = "WU-B-05-01, WU B 05-01",
    doi = "10.1140/epjc/s2005-02298-5",
    journal = "Eur. Phys. J. C",
    volume = "42",
    pages = "281--301",
    year = "2005"
}

@article{Goloskokov:2007nt,
    author = "Goloskokov, S. V. and Kroll, P.",
    title = "{The Role of the quark and gluon GPDs in hard vector-meson electroproduction}",
    eprint = "0708.3569",
    archivePrefix = "arXiv",
    primaryClass = "hep-ph",
    reportNumber = "WU-B-07-07, WU B 07-07",
    doi = "10.1140/epjc/s10052-007-0466-5",
    journal = "Eur. Phys. J. C",
    volume = "53",
    pages = "367--384",
    year = "2008"
}

@article{Grach:1983hd,
    author = "Grach, I. L. and Kondratyuk, L. A.",
    title = "{ELECTROMAGNETIC FORM-FACTOR OF DEUTERON IN RELATIVISTIC DYNAMICS. TWO NUCLEON AND SIX QUARK COMPONENTS}",
    reportNumber = "ITEP-59-1983",
    journal = "Sov. J. Nucl. Phys.",
    volume = "39",
    pages = "198",
    year = "1984"
}

@article{Gurjar:2024wpq,
    author = "Gurjar, Bheemsehan and Mondal, Chandan and Kaur, Satvir",
    title = {{{\ensuremath{\rho}}-meson spectroscopy and diffractive production using the holographic light-front Schr{\"o}dinger equation and the {\textquoteright}t Hooft equation}},
    eprint = "2401.13514",
    archivePrefix = "arXiv",
    primaryClass = "hep-ph",
    doi = "10.1103/PhysRevD.109.094017",
    journal = "Phys. Rev. D",
    volume = "109",
    number = "9",
    pages = "094017",
    year = "2024"
}

@article{Gurjar:2025kcp,
    author = "Gurjar, Bheemsehan and Mondal, Chandan and Kaur, Satvir",
    title = {{{\ensuremath{\phi}}-meson spectroscopy and diffractive production using two Schr{\"o}dinger-like equations on the light front}},
    eprint = "2501.11436",
    archivePrefix = "arXiv",
    primaryClass = "hep-ph",
    doi = "10.1103/PhysRevD.111.094002",
    journal = "Phys. Rev. D",
    volume = "111",
    number = "9",
    pages = "094002",
    year = "2025"
}

@article{H1rho1996,
    author = "Aid, S. and others",
    collaboration = "H1",
    title = "{Elastic electroproduction of $\rho^0$ and $J/\psi$ mesons at large Q$^2$ at HERA}",
    eprint = "hep-ex/9602007",
    archivePrefix = "arXiv",
    reportNumber = "DESY-96-023",
    doi = "10.1016/0550-3213(96)00192-7",
    journal = "Nucl. Phys. B",
    volume = "468",
    pages = "3--36",
    year = "1996",
    note = "[Erratum: Nucl.Phys.B 548, 639--639 (1999)]"
}

@article{H1rho2000,
    author = "Adloff, C. and others",
    collaboration = "H1",
    title = "{Elastic electroproduction of rho mesons at HERA}",
    eprint = "hep-ex/9902019",
    archivePrefix = "arXiv",
    reportNumber = "DESY-99-010",
    doi = "10.1007/s100520050703",
    journal = "Eur. Phys. J. C",
    volume = "13",
    pages = "371--396",
    year = "2000"
}

@article{H1rho2010,
    author = "Aaron, F. D. and others",
    collaboration = "H1",
    title = "{Diffractive Electroproduction of rho and phi Mesons at HERA}",
    eprint = "0910.5831",
    archivePrefix = "arXiv",
    primaryClass = "hep-ex",
    reportNumber = "DESY-09-093",
    doi = "10.1007/JHEP05(2010)032",
    journal = "JHEP",
    volume = "05",
    pages = "032",
    year = "2010"
}

@article{HERMESrho2009,
    author = "Airapetian, A. and others",
    collaboration = "HERMES",
    title = "{Spin Density Matrix Elements in Exclusive rho0 Electroproduction on H-1 and H-2 Targets at 27.5-GeV Beam Energy}",
    eprint = "0901.0701",
    archivePrefix = "arXiv",
    primaryClass = "hep-ex",
    reportNumber = "DESY-08-203",
    doi = "10.1140/epjc/s10052-009-1082-3",
    journal = "Eur. Phys. J. C",
    volume = "62",
    pages = "659--695",
    year = "2009"
}

@article{IancuItakuraMunier2004,
    author = "Iancu, E. and Itakura, K. and Munier, S.",
    title = "{Saturation and BFKL dynamics in the HERA data at small x}",
    eprint = "hep-ph/0310338",
    archivePrefix = "arXiv",
    reportNumber = "SACLAY-SPHT-T03-156, CPHT-RR-046-1003",
    doi = "10.1016/j.physletb.2004.02.040",
    journal = "Phys. Lett. B",
    volume = "590",
    pages = "199--208",
    year = "2004"
}

@article{Ivanov:2004ax,
    author = "Ivanov, I. P. and Nikolaev, N. N. and Savin, A. A.",
    title = "{Diffractive vector meson production at HERA: From soft to hard QCD}",
    eprint = "hep-ph/0501034",
    archivePrefix = "arXiv",
    reportNumber = "DESY-04-243",
    doi = "10.1134/S1063779606010011",
    journal = "Phys. Part. Nucl.",
    volume = "37",
    pages = "1--85",
    year = "2006"
}

@article{Ji:1998zi,
    author = "Ji, Chueng-Ryong and Choi, Ho-Meoyng",
    editor = "Jarczyk, Lucjan and Drozdz, S. and Magiera, A. and Machner, H.",
    title = "{Mixing angles and electromagnetic properties of ground state pseudoscalar and vector meson nonets in the light cone quark model}",
    journal = "Acta Phys. Polon. B",
    volume = "29",
    pages = "3363--3366",
    year = "1998"
}

@article{Kowalski:2003hm,
    author = "Kowalski, Henri and Teaney, Derek",
    title = "{An Impact parameter dipole saturation model}",
    eprint = "hep-ph/0304189",
    archivePrefix = "arXiv",
    doi = "10.1103/PhysRevD.68.114005",
    journal = "Phys. Rev. D",
    volume = "68",
    pages = "114005",
    year = "2003"
}

@article{KowalskiMotykaWatt2006,
    author = "Kowalski, H. and Motyka, L. and Watt, G.",
    title = "{Exclusive diffractive processes at HERA within the dipole picture}",
    eprint = "hep-ph/0606272",
    archivePrefix = "arXiv",
    reportNumber = "DESY-06-095",
    doi = "10.1103/PhysRevD.74.074016",
    journal = "Phys. Rev. D",
    volume = "74",
    pages = "074016",
    year = "2006"
}

@article{LepageBrodsky1980,
    author = "Lepage, G. Peter and Brodsky, Stanley J.",
    title = "{Exclusive Processes in Perturbative Quantum Chromodynamics}",
    reportNumber = "SLAC-PUB-2478",
    doi = "10.1103/PhysRevD.22.2157",
    journal = "Phys. Rev. D",
    volume = "22",
    pages = "2157",
    year = "1980"
}

@article{Martin:1996bp,
    author = "Martin, Alan D. and Ryskin, M. G. and Teubner, T.",
    title = "{The QCD description of diffractive rho meson electroproduction}",
    eprint = "hep-ph/9609448",
    archivePrefix = "arXiv",
    reportNumber = "DTP-96-64",
    doi = "10.1103/PhysRevD.55.4329",
    journal = "Phys. Rev. D",
    volume = "55",
    pages = "4329--4337",
    year = "1997"
}

@article{Melosh1974,
    author = "Melosh, H. J.",
    title = "{Quarks: Currents and constituents}",
    doi = "10.1103/PhysRevD.9.1095",
    journal = "Phys. Rev. D",
    volume = "9",
    pages = "1095",
    year = "1974"
}

@article{Mueller:1993rr,
    author = "Mueller, Alfred H.",
    title = "{Soft gluons in the infinite momentum wave function and the BFKL pomeron}",
    reportNumber = "SLAC-PUB-10047, CU-TP-609",
    doi = "10.1016/0550-3213(94)90116-3",
    journal = "Nucl. Phys. B",
    volume = "415",
    pages = "373--385",
    year = "1994"
}

@article{NMCrho1994,
    author = "Arneodo, M. and others",
    collaboration = "New Muon",
    title = "{Exclusive rho0 and phi muoproduction at large q**2}",
    reportNumber = "CERN-PPE-94-146",
    doi = "10.1016/0550-3213(94)90152-X",
    journal = "Nucl. Phys. B",
    volume = "429",
    pages = "503--529",
    year = "1994"
}

@article{Nemchik:1996cw,
    author = "Nemchik, J. and Nikolaev, Nikolai N. and Predazzi, E. and Zakharov, B. G.",
    title = "{Color dipole phenomenology of diffractive electroproduction of light vector mesons at HERA}",
    eprint = "hep-ph/9605231",
    archivePrefix = "arXiv",
    reportNumber = "DFTT-71-95, KFA-IKP-TH-95-24",
    doi = "10.1007/s002880050448",
    journal = "Z. Phys. C",
    volume = "75",
    pages = "71--87",
    year = "1997"
}

@article{Nikolaev:1990ja,
    author = "Nikolaev, Nikolai N. and Zakharov, B. G.",
    editor = "Khalatnikov, I. M. and Mineev, V. P.",
    title = "{Color transparency and scaling properties of nuclear shadowing in deep inelastic scattering}",
    reportNumber = "OUTP-90-23-P",
    doi = "10.1007/BF01483577",
    journal = "Z. Phys. C",
    volume = "49",
    pages = "607--618",
    year = "1991"
}

@article{PDG2024,
    author = "Navas, S. and others",
    collaboration = "Particle Data Group",
    title = "{Review of particle physics}",
    doi = "10.1103/PhysRevD.110.030001",
    journal = "Phys. Rev. D",
    volume = "110",
    number = "3",
    pages = "030001",
    year = "2024"
}

@article{Puhan:2025ujg,
    author = "Puhan, Satyajit and Sharma, Shubham and Kumar, Narinder and Dahiya, Harleen",
    title = "{Valence quark distribution of the rho meson using a light-front quark model}",
    eprint = "2511.10981",
    archivePrefix = "arXiv",
    primaryClass = "hep-ph",
    doi = "10.1103/xjkc-l5tr",
    journal = "Phys. Rev. D",
    volume = "113",
    number = "3",
    pages = "036030",
    year = "2026"
}

@article{QCDSF:2008tjq,
    author = "Gurtler, M. and others",
    editor = "Aubin, Christopher and Cohen, Saul and Dawson, Chris and Dudek, Jozef and Edwards, Robert and Joo, Balint and Lin, Huey-Wen and Orginos, Kostas and Richards, David and Thacker, Hank",
    collaboration = "QCDSF",
    title = "{Vector meson electromagnetic form factors}",
    doi = "10.22323/1.066.0051",
    journal = "PoS",
    volume = "LATTICE2008",
    pages = "051",
    year = "2008"
}

@article{Rezaeian2013,
    author = "Rezaeian, Amir H. and Siddikov, Marat and Van de Klundert, Merijn and Venugopalan, Raju",
    title = "{Analysis of combined HERA data in the Impact-Parameter dependent Saturation model}",
    eprint = "1212.2974",
    archivePrefix = "arXiv",
    primaryClass = "hep-ph",
    doi = "10.1103/PhysRevD.87.034002",
    journal = "Phys. Rev. D",
    volume = "87",
    number = "3",
    pages = "034002",
    year = "2013"
}

@article{Schilling:1973ag,
    author = "Schilling, K. and Wolf, G.",
    title = "{How to analyze vector meson production in inelastic lepton scattering}",
    reportNumber = "DESY-73-13",
    doi = "10.1016/0550-3213(73)90371-4",
    journal = "Nucl. Phys. B",
    volume = "61",
    pages = "381--413",
    year = "1973"
}

@article{Shi:2025mne,
    author = "Shi, Chao and Lu, Liming and Li, Jian-feng and Jia, Wenbao",
    title = "{Diffractive electroproduction of light vector particles: Leading Fock-state contribution in the presence of significant higher Fock-state effects}",
    eprint = "2508.01874",
    archivePrefix = "arXiv",
    primaryClass = "hep-ph",
    doi = "10.1103/42x8-bztl",
    journal = "Phys. Rev. D",
    volume = "112",
    number = "7",
    pages = "074027",
    year = "2025"
}

@article{Shultz:2015pfa,
    author = "Shultz, Christian J. and Dudek, Jozef J. and Edwards, Robert G.",
    title = "{Excited meson radiative transitions from lattice QCD using variationally optimized operators}",
    eprint = "1501.07457",
    archivePrefix = "arXiv",
    primaryClass = "hep-lat",
    reportNumber = "JLAB-THY-15-2004",
    doi = "10.1103/PhysRevD.91.114501",
    journal = "Phys. Rev. D",
    volume = "91",
    number = "11",
    pages = "114501",
    year = "2015"
}

@article{Shuvaev:1999ce,
    author = "Shuvaev, A. G. and Golec-Biernat, Krzysztof J. and Martin, Alan D. and Ryskin, M. G.",
    title = "{Off diagonal distributions fixed by diagonal partons at small x and xi}",
    eprint = "hep-ph/9902410",
    archivePrefix = "arXiv",
    reportNumber = "DTP-99-18",
    doi = "10.1103/PhysRevD.60.014015",
    journal = "Phys. Rev. D",
    volume = "60",
    pages = "014015",
    year = "1999"
}

@article{Tanisha:2025qda,
    author = "Tanisha and Puhan, Satyajit and Yadav, Anurag and Dahiya, Harleen",
    title = "{Valence quark distribution of light {\ensuremath{\rho}} and heavy J/{\ensuremath{\psi}} vector mesons in a light-cone quark model}",
    eprint = "2505.09213",
    archivePrefix = "arXiv",
    primaryClass = "hep-ph",
    doi = "10.1103/cmbw-vcds",
    journal = "Phys. Rev. D",
    volume = "112",
    number = "5",
    pages = "054035",
    year = "2025"
}

@article{ZEUSphi2005,
    author = "Chekanov, S. and others",
    collaboration = "ZEUS",
    title = "{Exclusive electroproduction of phi mesons at HERA}",
    eprint = "hep-ex/0504010",
    archivePrefix = "arXiv",
    reportNumber = "DESY-05-038",
    doi = "10.1016/j.nuclphysb.2005.04.009",
    journal = "Nucl. Phys. B",
    volume = "718",
    pages = "3--31",
    year = "2005"
}

@article{ZEUSrho1999,
    author = "Breitweg, J. and others",
    collaboration = "ZEUS",
    title = "{Exclusive electroproduction of $\rho^0$ and $J/\psi$ mesons at HERA}",
    eprint = "hep-ex/9808020",
    archivePrefix = "arXiv",
    reportNumber = "DESY-98-107, ANL-HEP-PR-98-95",
    doi = "10.1007/s100529901051",
    journal = "Eur. Phys. J. C",
    volume = "6",
    pages = "603--627",
    year = "1999"
}

@article{ZEUSrho2007,
    author = "Chekanov, S. and others",
    collaboration = "ZEUS",
    title = "{Exclusive rho0 production in deep inelastic scattering at HERA}",
    eprint = "0708.1478",
    archivePrefix = "arXiv",
    primaryClass = "hep-ex",
    reportNumber = "DESY-07-118",
    doi = "10.1186/1754-0410-1-6",
    journal = "PMC Phys. A",
    volume = "1",
    pages = "6",
    year = "2007"
}

@article{ZEUS:2016psi,
    author = "Abramowicz, H. and others",
    collaboration = "ZEUS",
    title = "{Measurement of the cross-section ratio $\sigma_{\psi(2S)}/\sigma_{J/\psi(1S)}$ in deep inelastic exclusive $ep$ scattering at HERA}",
    eprint = "1606.08652",
    archivePrefix = "arXiv",
    primaryClass = "hep-ex",
    reportNumber = "DESY-16-008",
    doi = "10.1016/j.nuclphysb.2016.06.010",
    journal = "Nucl. Phys. B",
    volume = "909",
    pages = "934--953",
    year = "2016"
}

@article{ZEUS:2022psi,
    author = "Abt, I. and others",
    collaboration = "ZEUS",
    title = "{Measurement of the cross-section ratio {\ensuremath{\sigma}}$_{\psi (2S)}$/{\ensuremath{\sigma}}$_{J/\psi (1S)}$ in exclusive photoproduction at HERA}",
    eprint = "2206.13343",
    archivePrefix = "arXiv",
    primaryClass = "hep-ex",
    reportNumber = "DESY-22-107",
    doi = "10.1007/JHEP12(2022)164",
    journal = "JHEP",
    volume = "12",
    pages = "164",
    year = "2022"
}

@article{H1:2002psi,
    author = "Adloff, C. and others",
    collaboration = "H1",
    title = "{Diffractive photoproduction of psi(2S) mesons at HERA}",
    eprint = "hep-ex/0205107",
    archivePrefix = "arXiv",
    reportNumber = "DESY-02-075",
    doi = "10.1016/S0370-2693(02)02275-X",
    journal = "Phys. Lett. B",
    volume = "541",
    pages = "251--264",
    year = "2002"
}

@article{H1:1999psi,
    author = "Adloff, C. and others",
    collaboration = "H1",
    title = "{Charmonium production in deep inelastic scattering at HERA}",
    eprint = "hep-ex/9903008",
    archivePrefix = "arXiv",
    reportNumber = "DESY-99-026",
    doi = "10.1007/s100520050762",
    journal = "Eur. Phys. J. C",
    volume = "10",
    pages = "373--393",
    year = "1999"
}

@article{H1:1998psi,
    author = "Adloff, C. and others",
    collaboration = "H1",
    title = "{Photoproduction of psi (2S) mesons at HERA}",
    eprint = "hep-ex/9711012",
    archivePrefix = "arXiv",
    reportNumber = "DESY-97-228",
    doi = "10.1016/S0370-2693(97)01529-3",
    journal = "Phys. Lett. B",
    volume = "421",
    pages = "385--394",
    year = "1998"
}

@article{H1:2005jpsi,
    author = "Aktas, A. and others",
    collaboration = "H1",
    title = "{Elastic J/psi production at HERA}",
    eprint = "hep-ex/0510016",
    archivePrefix = "arXiv",
    reportNumber = "DESY-05-161",
    doi = "10.1140/epjc/s2006-02519-5",
    journal = "Eur. Phys. J. C",
    volume = "46",
    pages = "585--603",
    year = "2006"
}

@article{H1:2013jpsi,
    author = "Alexa, C. and others",
    collaboration = "H1",
    title = "{Elastic and Proton-Dissociative Photoproduction of J/psi Mesons at HERA}",
    eprint = "1304.5162",
    archivePrefix = "arXiv",
    primaryClass = "hep-ex",
    reportNumber = "DESY-13-058",
    doi = "10.1140/epjc/s10052-013-2466-y",
    journal = "Eur. Phys. J. C",
    volume = "73",
    number = "6",
    pages = "2466",
    year = "2013"
}

@article{ZEUS:2004jpsi,
    author = "Chekanov, S. and others",
    collaboration = "ZEUS",
    title = "{Exclusive electroproduction of J/psi mesons at HERA}",
    eprint = "hep-ex/0404008",
    archivePrefix = "arXiv",
    reportNumber = "DESY-04-052",
    doi = "10.1016/j.nuclphysb.2004.06.034",
    journal = "Nucl. Phys. B",
    volume = "695",
    pages = "3--37",
    year = "2004"
}

@article{LHCb:2024psi,
    author = "Aaij, Roel and others",
    collaboration = "LHCb",
    title = "{Measurement of exclusive $J/\psi$ and $\psi(2S)$ production at $\sqrt{s}=13$ TeV}",
    eprint = "2409.03496",
    archivePrefix = "arXiv",
    primaryClass = "hep-ex",
    reportNumber = "LHCb-PAPER-2024-012, CERN-EP-2024-213",
    doi = "10.21468/SciPostPhys.18.2.071",
    journal = "SciPost Phys.",
    volume = "18",
    number = "2",
    pages = "071",
    year = "2025"
}

@article{Jamal:2026heavy,
    author = "Jamal, Mohammad Yousuf and Islam, Ajaharul and Bandyopadhyay, Aritra and Das, Santosh K.",
    title = "{Diffractive production of heavy quarkonia at the Electron Ion Collider}",
    eprint = "2606.18854",
    archivePrefix = "arXiv",
    primaryClass = "hep-ph",
    doi = "10.1103/bplx-hsh5",
    journal = "Phys. Rev. D",
    volume = "114",
    number = "3",
    pages = "L031504",
    year = "2026"
}

@article{Dhiman:2019ddr,
    author = "Dhiman, Nisha and Dahiya, Harleen and Ji, Chueng-Ryong and Choi, Ho-Meoyng",
    title = "{Twist-2 Pseudoscalar and Vector Meson Distribution Amplitudes in Light-Front Quark Model with Exponential-type Confining Potential}",
    eprint = "1902.09160",
    archivePrefix = "arXiv",
    primaryClass = "hep-ph",
    doi = "10.1103/PhysRevD.100.014026",
    journal = "Phys. Rev. D",
    volume = "100",
    number = "1",
    pages = "014026",
    year = "2019"
}

@article{Ridwan:2025,
    author = "Ridwan, Muhammad and Arifi, Ahmad Jafar and Mart, Terry",
    title = "{Self-consistent M1 radiative transitions of excited Bc and heavy quarkonia with different polarizations in the light-front quark model}",
    eprint = "2409.13172",
    archivePrefix = "arXiv",
    primaryClass = "hep-ph",
    doi = "10.1103/PhysRevD.111.016011",
    journal = "Phys. Rev. D",
    volume = "111",
    number = "1",
    pages = "016011",
    year = "2025"
}

@article{Arifi:2022pal,
    author = "Arifi, Ahmad Jafar and Choi, Ho-Meoyng and Ji, Chueng-Ryong and Oh, Yongseok",
    title = "{Mixing effects on 1S and 2S state heavy mesons in the light-front quark model}",
    eprint = "2205.04075",
    archivePrefix = "arXiv",
    primaryClass = "hep-ph",
    doi = "10.1103/PhysRevD.106.014009",
    journal = "Phys. Rev. D",
    volume = "106",
    number = "1",
    pages = "014009",
    year = "2022"
}

@article{Nemchik:1997xb,
    author = "Nemchik, J. and Nikolaev, Nikolai N. and Predazzi, E. and Zakharov, B. G. and Zoller, V. R.",
    title = "{The Diffraction cone for exclusive vector meson production in deep inelastic scattering}",
    eprint = "hep-ph/9712469",
    archivePrefix = "arXiv",
    reportNumber = "FZ-IKP-97-23",
    doi = "10.1134/1.558573",
    journal = "J. Exp. Theor. Phys.",
    volume = "86",
    pages = "1054--1073",
    year = "1998"
}

@article{Hoyer:1999xe,
    author = "Hoyer, Paul and Peigne, Stephane",
    title = "{$J^\prime / \psi^\prime$ to $J/\psi$ ratio in diffractive photoproduction}",
    eprint = "hep-ph/9909519",
    archivePrefix = "arXiv",
    reportNumber = "NORDITA-1999-59-HE, LAPTH-749-99",
    doi = "10.1103/PhysRevD.61.031501",
    journal = "Phys. Rev. D",
    volume = "61",
    pages = "031501",
    year = "2000"
}

@article{Suzuki:2000az,
    author = "Suzuki, K. and Hayashigaki, A. and Itakura, K. and Alam, J. and Hatsuda, T.",
    title = "{Validity of the color dipole approximation for diffractive production of heavy quarkonium}",
    eprint = "hep-ph/0005250",
    archivePrefix = "arXiv",
    doi = "10.1103/PhysRevD.62.031501",
    journal = "Phys. Rev. D",
    volume = "62",
    pages = "031501",
    year = "2000"
}

@article{Frankfurt:1997fj,
    author = "Frankfurt, Leonid and Koepf, Werner and Strikman, Mark",
    title = "{Diffractive heavy quarkonium photoproduction and electroproduction in QCD}",
    eprint = "hep-ph/9702216",
    archivePrefix = "arXiv",
    reportNumber = "OSU-97-0201, DESY-97-028",
    doi = "10.1103/PhysRevD.57.512",
    journal = "Phys. Rev. D",
    volume = "57",
    pages = "512--526",
    year = "1998"
}

@article{Cepila:2019skb,
    author = "Cepila, Jan and Nemchik, Jan and Krelina, Michal and Pasechnik, Roman",
    title = "{Theoretical uncertainties in exclusive electroproduction of S-wave heavy quarkonia}",
    eprint = "1901.02664",
    archivePrefix = "arXiv",
    primaryClass = "hep-ph",
    doi = "10.1140/epjc/s10052-019-7016-9",
    journal = "Eur. Phys. J. C",
    volume = "79",
    number = "6",
    pages = "495",
    year = "2019"
}

@article{Krelina:2019gee,
    author = "Krelina, M. and Goncalves, V. P. and Cepila, J.",
    title = "{Coherent and incoherent vector meson electroproduction in the future electron-ion colliders: the hot-spot predictions}",
    eprint = "1905.06759",
    archivePrefix = "arXiv",
    primaryClass = "hep-ph",
    doi = "10.1016/j.nuclphysa.2019.06.009",
    journal = "Nucl. Phys. A",
    volume = "989",
    pages = "187--200",
    year = "2019"
}

@article{Krelina:2019egg,
    author = "Krelina, Michal and Nemchik, Jan and Pasechnik, Roman",
    title = "{$D$-wave effects in diffractive electroproduction of heavy quarkonia from the photon-like $V\rightarrow Q{\bar{Q}}$ transition}",
    eprint = "1909.12770",
    archivePrefix = "arXiv",
    primaryClass = "hep-ph",
    doi = "10.1140/epjc/s10052-020-7678-3",
    journal = "Eur. Phys. J. C",
    volume = "80",
    number = "2",
    pages = "92",
    year = "2020"
}

@article{Henkels:2020kju,
    author = "Henkels, Cheryl and de Oliveira, Emmanuel G. and Pasechnik, Roman and Trebien, Haimon",
    title = "{Exclusive photoproduction of excited quarkonia in ultraperipheral collisions}",
    eprint = "2004.00607",
    archivePrefix = "arXiv",
    primaryClass = "hep-ph",
    doi = "10.1103/PhysRevD.102.014024",
    journal = "Phys. Rev. D",
    volume = "102",
    number = "1",
    pages = "014024",
    year = "2020"
}

@article{Mantysaari:2021ryb,
    author = {M{\"a}ntysaari, Heikki and Penttala, Jani},
    title = "{Exclusive heavy vector meson production at next-to-leading order in the dipole picture}",
    eprint = "2104.02349",
    archivePrefix = "arXiv",
    primaryClass = "hep-ph",
    doi = "10.1016/j.physletb.2021.136723",
    journal = "Phys. Lett. B",
    volume = "823",
    pages = "136723",
    year = "2021"
}

@article{Peredo:2023oym,
    author = "Peredo, Marco Alcazar and Hentschinski, Martin",
    title = "{Ratio of J/{\ensuremath{\Psi}} and {\ensuremath{\Psi}}(2s) exclusive photoproduction cross sections as an indicator for the presence of nonlinear QCD evolution}",
    eprint = "2308.15430",
    archivePrefix = "arXiv",
    primaryClass = "hep-ph",
    doi = "10.1103/PhysRevD.109.014032",
    journal = "Phys. Rev. D",
    volume = "109",
    number = "1",
    pages = "014032",
    year = "2024"
}

@article{Mantysaari:2022kdm,
    author = {M{\"a}ntysaari, Heikki and Penttala, Jani},
    title = "{Exclusive production of light vector mesons at next-to-leading order in the dipole picture}",
    eprint = "2203.16911",
    archivePrefix = "arXiv",
    primaryClass = "hep-ph",
    doi = "10.1103/PhysRevD.105.114038",
    journal = "Phys. Rev. D",
    volume = "105",
    number = "11",
    pages = "114038",
    year = "2022"
}

@article{Lappi:2020ufv,
    author = {Lappi, Tuomas and M{\"a}ntysaari, Heikki and Penttala, Jani},
    title = "{Relativistic corrections to the vector meson light front wave function}",
    eprint = "2006.02830",
    archivePrefix = "arXiv",
    primaryClass = "hep-ph",
    doi = "10.1103/PhysRevD.102.054020",
    journal = "Phys. Rev. D",
    volume = "102",
    number = "5",
    pages = "054020",
    year = "2020"
}

@article{Dudek:2006ej,
    author = "Dudek, Jozef J. and Edwards, Robert G. and Richards, David G.",
    title = "{Radiative transitions in charmonium from lattice QCD}",
    eprint = "hep-ph/0601137",
    archivePrefix = "arXiv",
    reportNumber = "JLAB-THY-06-457",
    doi = "10.1103/PhysRevD.73.074507",
    journal = "Phys. Rev. D",
    volume = "73",
    pages = "074507",
    year = "2006"
}

@article{Delaney:2023fsx,
    author = "Delaney, James and Thomas, Christopher E. and Ryan, Sin{\'e}ad M.",
    collaboration = "Hadron Spectrum",
    title = "{Radiative transitions in charmonium from lattice QCD}",
    eprint = "2301.08213",
    archivePrefix = "arXiv",
    primaryClass = "hep-lat",
    doi = "10.1007/JHEP05(2024)230",
    journal = "JHEP",
    volume = "05",
    pages = "230",
    year = "2024"
}

@article{Li:2018jpsi,
    author = "Li, Ning and Wu, Ya-Jie and Gao, Tan-Biao",
    title = "{Lattice study of $J/\Psi$ form factors with twisted boundary conditions}",
    doi = "10.1140/epja/i2018-12627-x",
    journal = "Eur. Phys. J. A",
    volume = "54",
    number = "11",
    pages = "197",
    year = "2018"
}

@article{H1ZEUS:2015HERA,
    author = "Abramowicz, H. and others",
    collaboration = "H1, ZEUS",
    title = "{Combination of measurements of inclusive deep inelastic ${e^{\pm }p}$ scattering cross sections and QCD analysis of HERA data}",
    eprint = "1506.06042",
    archivePrefix = "arXiv",
    primaryClass = "hep-ex",
    reportNumber = "DESY-15-039",
    doi = "10.1140/epjc/s10052-015-3710-4",
    journal = "Eur. Phys. J. C",
    volume = "75",
    number = "12",
    pages = "580",
    year = "2015"
}

@article{Soyez:2007kg,
    author = "Soyez, G.",
    title = "{Saturation QCD predictions with heavy quarks at HERA}",
    eprint = "0705.3672",
    archivePrefix = "arXiv",
    primaryClass = "hep-ph",
    doi = "10.1016/j.physletb.2007.07.076",
    journal = "Phys. Lett. B",
    volume = "655",
    pages = "32--38",
    year = "2007"
}

@article{Watt:2007nr,
    author = "Watt, G. and Kowalski, H.",
    title = "{Impact parameter dependent colour glass condensate dipole model}",
    eprint = "0712.2670",
    archivePrefix = "arXiv",
    primaryClass = "hep-ph",
    doi = "10.1103/PhysRevD.78.014016",
    journal = "Phys. Rev. D",
    volume = "78",
    pages = "014016",
    year = "2008"
}

@article{Donnachie:1992ny,
    author = "Donnachie, A. and Landshoff, P. V.",
    title = "{Total cross-sections}",
    eprint = "hep-ph/9209205",
    archivePrefix = "arXiv",
    reportNumber = "CERN-TH-6635-92",
    doi = "10.1016/0370-2693(92)90832-O",
    journal = "Phys. Lett. B",
    volume = "296",
    pages = "227--232",
    year = "1992"
}

@article{EicC:2021,
    author = "Anderle, Daniele P. and others",
    title = "{Electron-ion collider in China}",
    eprint = "2102.09222",
    archivePrefix = "arXiv",
    primaryClass = "nucl-ex",
    reportNumber = "Frontiers of Physics, Volume 16 Issue (6):64701, 2021",
    doi = "10.1007/s11467-021-1062-0",
    journal = "Front. Phys. (Beijing)",
    volume = "16",
    number = "6",
    pages = "64701",
    year = "2021"
}

@article{LHeC:2020,
    author = "Agostini, P. and others",
    collaboration = "LHeC, FCC-he Study Group",
    title = "{The Large Hadron{\textendash}Electron Collider at the HL-LHC}",
    eprint = "2007.14491",
    archivePrefix = "arXiv",
    primaryClass = "hep-ex",
    reportNumber = "CERN-ACC-Note-2020-0002, JLAB-ACP-20-3180",
    doi = "10.1088/1361-6471/abf3ba",
    journal = "J. Phys. G",
    volume = "48",
    number = "11",
    pages = "110501",
    year = "2021"
}

@article{Chung:1988my,
    author = "Chung, P. L. and Polyzou, W. N. and Coester, F. and Keister, B. D.",
    title = "{Hamiltonian Light Front Dynamics of Elastic electron Deuteron Scattering}",
    doi = "10.1103/PhysRevC.37.2000",
    journal = "Phys. Rev. C",
    volume = "37",
    pages = "2000--2015",
    year = "1988"
}

@article{Frankfurt:1993ut,
    author = "Frankfurt, L. L. and Strikman, M. and Frederico, T.",
    title = "{Deuteron form-factors in the light cone quantum mechanics 'good' component approach}",
    doi = "10.1103/PhysRevC.48.2182",
    journal = "Phys. Rev. C",
    volume = "48",
    pages = "2182--2189",
    year = "1993"
}

@article{E665:1996mob,
    author = "Adams, M. R. and others",
    collaboration = "E665",
    title = "{Proton and Deuteron Structure Functions in Muon Scattering at 470 GeV}",
    reportNumber = "FERMILAB-PUB-95-396-E",
    doi = "10.1103/PhysRevD.54.3006",
    journal = "Phys. Rev. D",
    volume = "54",
    pages = "3006--3056",
    year = "1996"
}

@article{NewMuon:1996fwh,
    author = "Arneodo, M. and others",
    collaboration = "New Muon",
    title = "{Measurement of the proton and deuteron structure functions, F2(p) and F2(d), and of the ratio sigma-L / sigma-T}",
    eprint = "hep-ph/9610231",
    archivePrefix = "arXiv",
    doi = "10.1016/S0550-3213(96)00538-X",
    journal = "Nucl. Phys. B",
    volume = "483",
    pages = "3--43",
    year = "1997"
}

@article{H1:2018flt,
    author = "Abramowicz, H. and others",
    collaboration = "H1, ZEUS",
    title = "{Combination and QCD analysis of charm and beauty production cross-section measurements in deep inelastic $ep$ scattering at HERA}",
    eprint = "1804.01019",
    archivePrefix = "arXiv",
    primaryClass = "hep-ex",
    reportNumber = "DESY 18-037, DESY-18-037",
    doi = "10.1140/epjc/s10052-018-5848-3",
    journal = "Eur. Phys. J. C",
    volume = "78",
    number = "6",
    pages = "473",
    year = "2018"
}

@article{H1:2013ktq,
    author = "Andreev, V. and others",
    collaboration = "H1",
    title = "{Measurement of inclusive $e p$ cross sections at high $Q^2$ at $\sqrt s =$ 225 and 252 GeV and of the longitudinal proton structure function $F_L$ at HERA}",
    eprint = "1312.4821",
    archivePrefix = "arXiv",
    primaryClass = "hep-ex",
    reportNumber = "DESY-13-211",
    doi = "10.1140/epjc/s10052-014-2814-6",
    journal = "Eur. Phys. J. C",
    volume = "74",
    number = "4",
    pages = "2814",
    year = "2014"
}

@article{Virtanen:2019joe,
    author = "Virtanen, Pauli and others",
    title = "{SciPy 1.0--Fundamental Algorithms for Scientific Computing in Python}",
    eprint = "1907.10121",
    archivePrefix = "arXiv",
    primaryClass = "cs.MS",
    doi = "10.1038/s41592-019-0686-2",
    journal = "Nature Meth.",
    volume = "17",
    pages = "261",
    year = "2020"
}

@article{Golec-Biernat:2017lfv,
    author = "Golec-Biernat, Krzysztof and Sapeta, Sebastian",
    title = "{Saturation model of DIS : an update}",
    eprint = "1711.11360",
    archivePrefix = "arXiv",
    primaryClass = "hep-ph",
    reportNumber = "IFJPAN-IV-2017-27",
    doi = "10.1007/JHEP03(2018)102",
    journal = "JHEP",
    volume = "03",
    pages = "102",
    year = "2018"
}

\end{document}